\documentclass[final,3p,times,12pt]{elsarticle}
\usepackage{amssymb}
\usepackage{graphicx,type 1cm,eso-pic,color}

\makeatletter
\def\ps@pprintTitle{%
  \let\@oddhead\@empty
  \let\@evenhead\@empty
  \let\@oddfoot\@empty
  \let\@evenfoot\@oddfoot
}
\makeatother

\begin{document}
\begin{frontmatter}

\title{
The {\sf ArgoNeuT} Detector\\ in the NuMI Low-Energy Beam Line at Fermilab}

\author[Yale]{C.~Anderson}
\author[gs]{M.~Antonello}
\author[FNAL]{B.~Baller}
\author[ksu]{T.~Bolton}
\author[msu]{C.~Bromberg}
\author[laq,Yale]{F.~Cavanna}
\author[Yale]{E.~Church}
\author[msu]{D.~Edmunds}
\author[bern]{A.~Ereditato}
\author[ksu]{S.~Farooq}
\author[Yale]{B.~Fleming}
\author[FNAL]{H.~Greenlee}
\author[Yale]{R.~Guenette}
\author[bern]{S.~Haug}
\author[ksu]{G.~Horton-Smith}
\author[FNAL]{C.~James}
\author[Yale]{E.~Klein}
\author[tex]{K.~Lang}
\author[FNAL]{A.~Lathrop}
\author[msu]{P.~Laurens}
\author[Yale]{S.~Linden}
\author[ksu]{D.~McKee}
\author[tex]{R.~Mehdiyev}
\author[msu]{B.~Page}
\author[gs,Yale]{O.~Palamara}
\author[Yale]{K.~Partyka}
\author[FNAL]{S.~Pordes}
\author[FNAL]{G.~Rameika}
\author[FNAL]{B.~Rebel}
\author[bern]{B.~Rossi}
\author[FNAL]{R.~Sanders}
\author[Syracuse,FNAL]{M.~Soderberg}
\author[Yale]{J.~Spitz}
\author[Yale]{A.M.~Szelc}
\author[bern]{M.~Weber}
\author[FNAL]{T.~Yang}
\author[Yale]{T.~Wongjirad}
\author[FNAL]{G.~Zeller}

\address[Yale]{Yale University, New Haven, Connecticut - USA}
\address[gs]{INFN - Laboratori Nazionali del Gran Sasso, Assergi - Italy}
\address[FNAL] {Fermi National Accelerator Laboratory, Chicago, Illinois - USA}
\address[ksu]{Kansas State University, Manhattan, Kansas - USA}
\address[msu]{Michigan State University, East Lansing, Michigan - USA}
\address[laq]{University of L'Aquila,  L'Aquila - Italy}
\address[bern]{University of Bern, Bern - Switzerland}
\address[tex]{University of Texas at Austin,  Austin, Texas - USA}
\address[Syracuse]{Syracuse University, Syracuse, New York - USA}

\begin{abstract}
The {\sf ArgoNeuT} liquid argon time projection chamber has collected thousands of neutrino and anti-neutrino events during an extended run period in the NuMI beam-line at Fermilab. This paper focuses on the main aspects of the detector layout and related technical features, 
including the cryogenic equipment, time projection chamber, read-out electronics, and off-line data treatment.
The detector commissioning phase, physics run, and first neutrino event displays are also reported. The characterization of the main working parameters of the detector during data-taking, the ionization electron drift velocity and lifetime in liquid argon,  as obtained from through-going muon data complete the present report.
\end{abstract}


\end{frontmatter}

\date{\today}




\section{Introduction}
\label{Introduction}
Liquid argon time projection chambers (LArTPCs) offer a unique combination of fine-grained tracking, precise calorimetry, and scalability to very large sensitive volumes. This makes this detection technology ideally suited for the study of neutrino interactions and searches for rare phenomena.  The LArTPC concept was proposed in the late seventies, followed by a long history of technological development in Europe~\cite{CR,T600-PVtest} leading up to the realization of a first 
detector of significant size for underground physics application (ICARUS T600)~\cite{T600-LNGS}. 

Currently, there is worldwide interest in utilizing this technology, with the goal of deploying multi-kiloton LArTPCs in far-detector locations as part of long-baseline neutrino oscillation and proton decay search 
programs~\cite{LBNE,LAGUNA,LArJap}. 

In the US, along the path of its phased program towards the construction of a massive LArTPC detector for LBNE \cite{LBNE}, 
as a first step the {\sf ArgoNeuT} detector was built and operated on neutrino beams at Fermilab 
for $\nu$-Ar cross section measurements, followed by 
the MicroBooNE Experiment~\cite{MicroBooNE} now under construction. 
MicroBooNE, with approximately 100-ton of liquid argon TPC,
will investigate on sterile neutrino oscillations at Fermilab.

{\sf ArgoNeuT}, a NSF/DOE project at Fermilab (T962), is the first LArTPC operated in a ``low-energy"  neutrino beam (neutrino 
energies in the 0.5-10.0~GeV range). These energies are most relevant for long-baseline
neutrino oscillation searches as the oscillation probability is maximal in the few GeV
region, assuming typical values of $\theta_{13}\sim 8^{\circ}$, $\theta_{23}\sim 45^{\circ}$,
$\Delta m_{13}^{2}\sim 2.4\times10^{-3}~\mathrm{eV}^{2}$, and the current baseline length option of about $1000~\mathrm{km}$. 

The {\sf ArgoNeuT} experiment's operations began with a cosmic ray commissioning run on the surface in Summer 2008 and a cosmic ray and beam-induced neutrino commissioning run underground in Spring 2009. The surface run took place at the Proton Assembly Building at Fermilab and the underground runs (commissioning and physics) were in the MINOS near detector hall, about 100 metres below ground. The first neutrino candidate was recorded on May 27, 2009.

\begin{figure}[tb]
\begin{centering}
\begin{tabular}{c}
\includegraphics[height=2.4in]{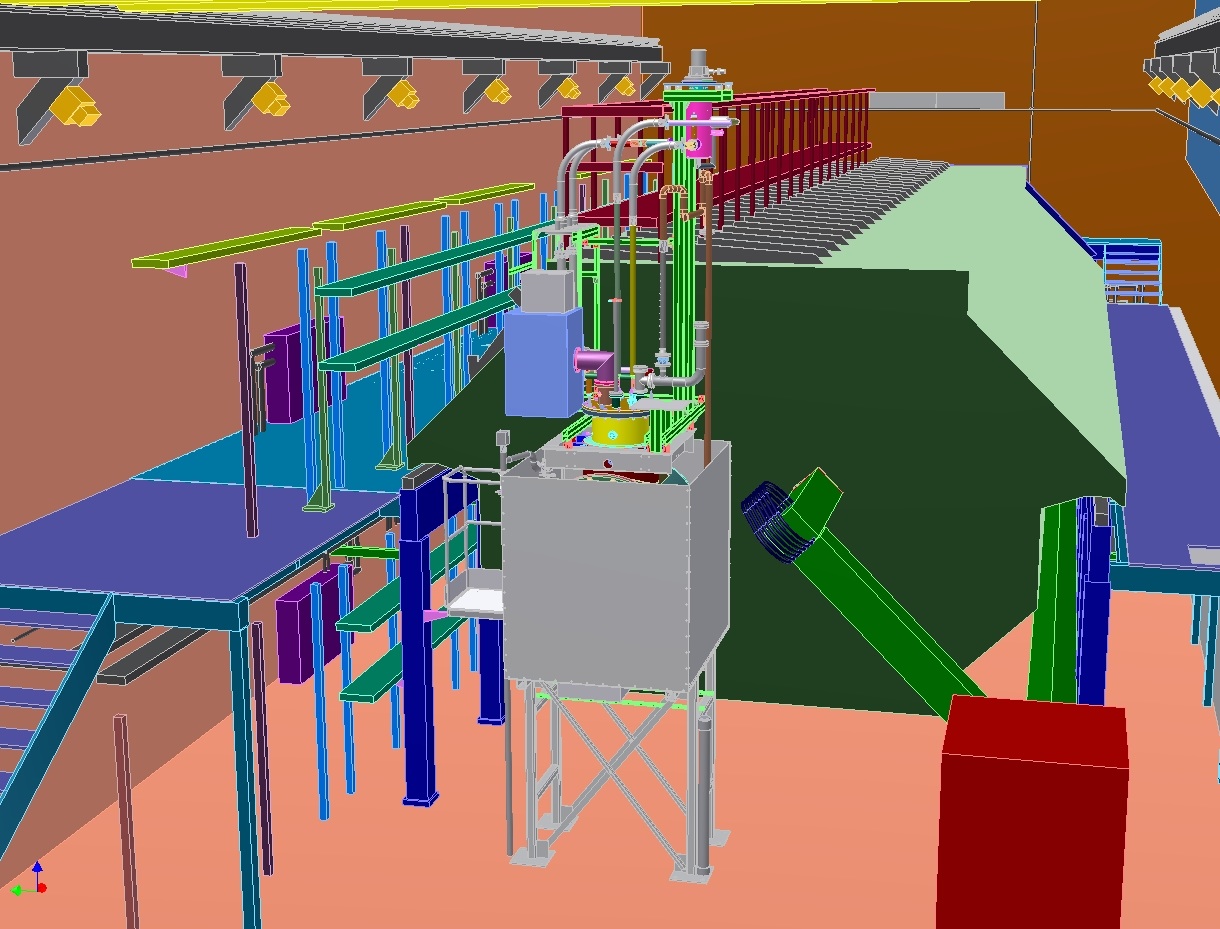} 
\includegraphics[height=2.4in]{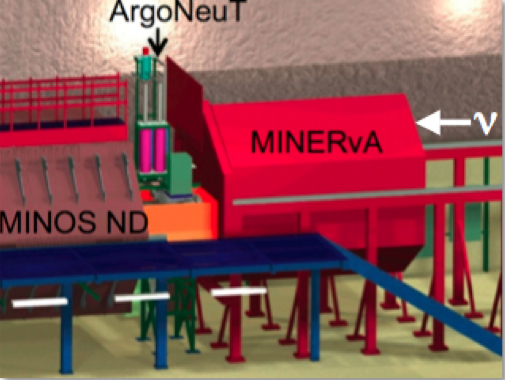} 
\end{tabular}
\caption{Rendering of the MINOS-ND hall. [Left] {\sf ArgoNeuT}, inside the gray box upstream of MINOS-ND, is positioned approximately at the center of the NuMI beam. [Right] In the location just upstream of {\sf ArgoNeuT} the MINER$\nu$A experiment was installed over a period of months during the {\sf ArgoNeuT} physics run.}
\label{minos_innearhall}
\end{centering}
\end{figure}

The physics run began in September 2009. During this time, {\sf ArgoNeuT} was located just upstream of the MINOS Near Detector (MINOS-ND) ~\cite{MINOS-ND-tech}, with the TPC centered 26~cm below the center of the NuMI on-axis beam\cite{NuMI-beam}. The MINOS-ND was used as a range stack to measure uncontained long-track muons from charged current neutrino interactions in the
{\sf ArgoNeuT} active volume. A rendering of {\sf ArgoNeuT}'s location in the MINOS-ND hall can be seen in figure~\ref{minos_innearhall}.  
The run lasted until late February 2010 and consisted of about two weeks of neutrino-mode running and four-and-a-half months of anti-neutrino mode running. Run operations were largely stable and shift-free over more than five months time period. 

\section{The {\sf ArgoNeuT} Detector}
\label{Detector}
\begin{figure}[tb]
\begin{centering}
\begin{tabular}{c c}
\includegraphics[height=3.3in]{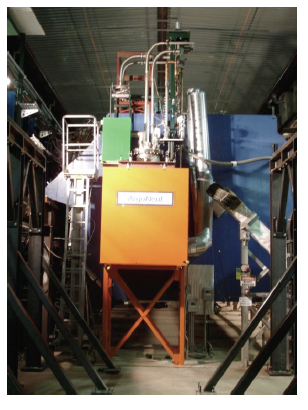} 
&
\includegraphics[height=3.3in]{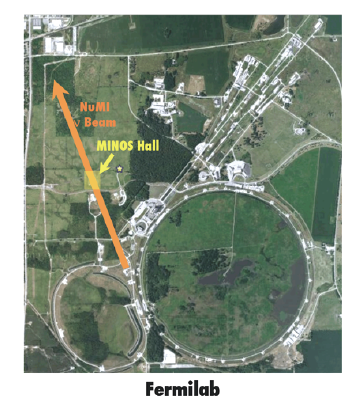}
\end{tabular}
\vspace{-0cm}
\caption{[Left] The fully instrumented {\sf ArgoNeuT} detector in the neutrino beam-line. The cryostat is placed into the orange box to contain spilled liquid argon in the (remote) case of a gross leak. MINOS-ND is visible behind {\sf ArgoNeuT}.
[Right] Aerial view of Fermilab. The MINOS-ND site is located about 1~km from the NuMI target station, about 100 m below ground level. The beam is produced by the Main Injector (the bottom left of the image, adjacent to the larger TeVatron ring).}
\label{detectorpic}
\end{centering}
\end{figure}

{\sf ArgoNeuT} consists of a vacuum insulated cryostat for ultra-pure liquid argon (LAr) containment, in which is mounted a time projection chamber with its field-shaping system.  The anode of the TPC, opposite to the cathode at one side of the detector volume, consists of three planes of parallel wires with different orientations. The TPC is operated at uniform electric field between cathode and anode.\\
Feed-throughs on an exit flange on top of the cryostat provide the electrical connection of the wires to the read-out electronics.
A front-end integrating preamplifier, followed by high- and low-pass filters,  make up the analog readout electronics of the {\sf ArgoNeuT} experiment. Wire signals are digitized and recorded at each NuMI beam spill trigger. \\
{\sf ArgoNeuT} uses a closed-loop recirculation and purification system to maintain a clean and constant volume of liquid argon, while continually purifying.  A  cryocooler is mounted above the cryostat and is used to re-condense boil-off vapor from the liquid volume.  Purification is achieved by directing the re-condensed liquid into a filter that removes electronegative impurities.  \\
A picture of {\sf ArgoNeuT} and its location in the MINOS-ND hall can be seen in figure~\ref{detectorpic} and details of the main components of the experimental set-up are given in the following subsections.

\subsection{Cryostat and Cryocooler}
\label{CryCry}
{\sf ArgoNeuT}'s liquid argon is contained by a double-wall, vacuum jacketed and super-insulated cryostat,
custom designed and manufactured from Type 304/304L-grade stainless steel.
The shape is cylindrical with convex end-caps at both ends. The main axis of the cryostat is horizontal and oriented parallel to the beam.  The inner vessel is $76.2$ cm (30'') in diameter ($\o$) and 130 cm in length corresponding to a LAr volume of about 550 L (0.76 t).  The overall external dimensions of the cryostat are 106.7 cm (42'') in diameter and 163 cm in length. 
Access to the internal volume, e.g. for detector installation, is possible by opening the end-caps (inner and outer vessels) at one end of the cryostat. These are 32-bolt flanged end-caps with a double Helicoflex\textsuperscript{\textregistered} sealing with guard vacuum for the inner vessel at LAr temperature and with Viton\textsuperscript{\textregistered}  O-ring sealing for the outer vessel at room temperature.  \\
\begin{figure}[h]
\begin{centering}
\begin{tabular}{c c}
\includegraphics[height=2.6in]{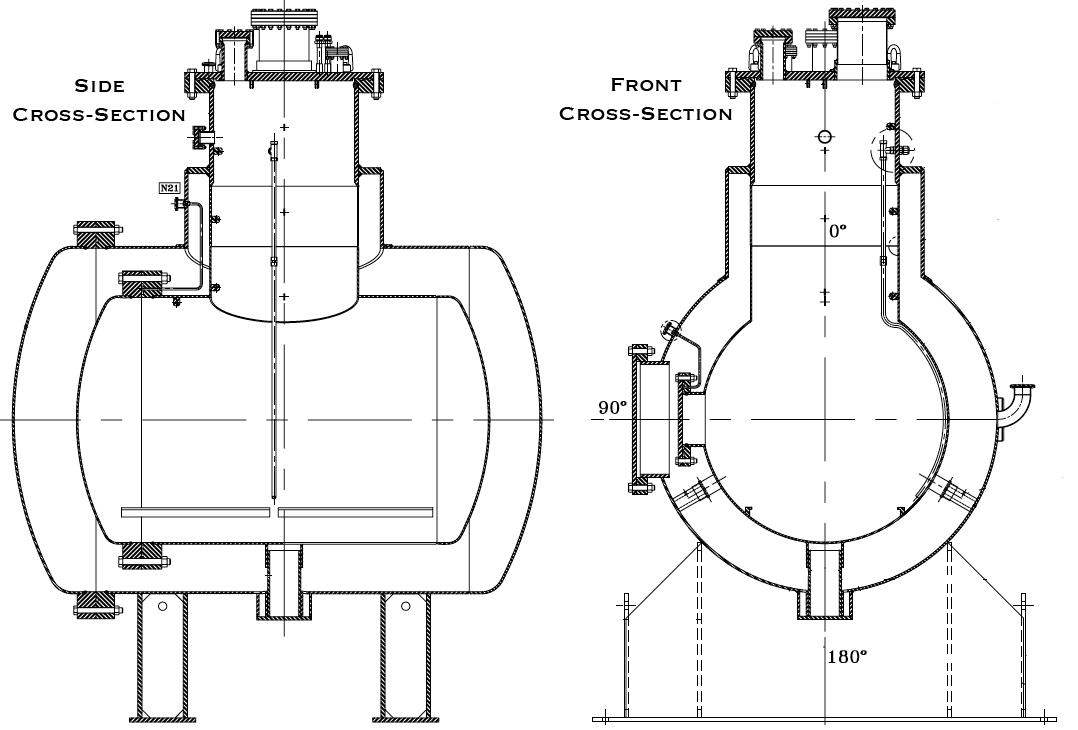} 
&
\includegraphics[height=4.0in]{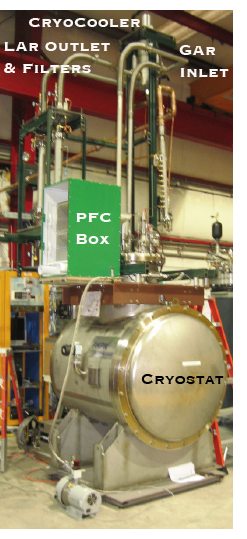} 
\end{tabular}
\caption{[Left] Side and Front Cross-Section views of the {\sf ArgoNeuT} cryostat. The inner and outer vessels, the chimney on the top and the removable end-caps at one side are visible in the drawings. [Right] Picture taken during assembly with details of the cooling and recirculation system (the 4-pipe pathways of the recirculation circuit and the cold head of the cryocooler located inside a vacuum-insulated containment vessel).}
\label{cryo_setup}
\end{centering}
\end{figure}
The cryostat has a wide neck consisting of a partially vacuum jacketed chimney formed by two coaxial stainless-steel, straight-joints assembly (internal diameter 46 cm, 60 cm in height). The chimney is located on the top of the cryostat at its mid length and serves as access path for signal cables from the TPC and from the internal instrumentation, as well as for the outlet/inlet pipes for the gas/liquid Ar recirculation and for the high voltage (HV) feed-through. A series of radiation baffles designed to reduce the effects of cryogenic radiation are also placed inside the chimney.
The neck is closed at the top by a 45.7 cm (18'') flange (double Helicoflex\textsuperscript{\textregistered} sealing with guard vacuum) 
with several ports equipped with electrical feed-throughs or valves. 
The inner vessel is welded to the outer vessel in the chimney assembly, and hangs from this weld.
Inside the chimney Ar in gas phase (GAr) is at equilibrium pressure with the liquid inside the cryostat main body.
\begin{table}[h]
\centering
\begin{tabular}{|c|c|}
	\hline
Liquid Argon volume (mass) & 550 liters  (0.77 t) \\ \hline
Inner Vessel Dimensions & $\o$=30'', $l$=130 cm \\ \hline
Outer Vessel Dimensions & $\o$=42'', $l$=163 cm \\ \hline
Insulation & Vacuum Jacket ($10^{-4}$ mbar) with SuperInsulation \\ \hline
Total Heat Load & $\approx$ 120  W \\ \hline
Cooling & CryoCooler (330 W cooling capacity) \\ \hline
Ar Recondensation & LAr Flow Rate: $\approx$ 3 lt/hr \\ \hline
P, T (set point) & GAr P=2 psig, LAr T=88.4 K  \\ \hline
\hline
\end{tabular}
\caption{{\sf ArgoNeuT} cryostat and cryogenic system main specifications.}
\label{cryo}
\end{table}

The pressure in the vacuum jacket is kept in the low $10^{-4}$~mbar range for the entire length of the physics run and, also due to the multilayer  radiation barrier (MLI, including 20 layers of aluminized mylar) in the jacket,
the overall residual  heat load of the system  is estimated in the range of 120-160 W. \\
To keep the argon inside the cryostat in liquid phase at constant temperature (around 88 K), the cryogenic system is based on a commercial single stage cryocooler with high cooling capacity, in excess of 330 W at LAr temperature.
The compressor package and the cold head expander are separated, and connected with flexible lines. The water cooled compressor package supplies  compressed helium for the cold head inside of which the helium expands to create cryogenic temperatures. \\
A copper finned heat exchanger, bolted to the cold head bottom, is sealed inside a small vacuum insulated vessel connected to the cryostat by a four-pipe vertical pathway (about 2 m rising from the top flange of the chimney). Boil-off argon gas from the surface of the liquid volume travels vertically through one pipe (the GAr inlet) and is re-condensed inside the heat exchanger vessel. The resulting liquid is then forced through one of the other three pipes (LAr outlet) on its return trip back to the liquid volume at the bottom of the cryostat. Two of these three passes are through argon filters and one is through a bypass pipe. The filters and pathways are also vacuum insulated. During operation the bypass is closed and the argon re-condensation and purification closed-loop cycle is enabled through one filter.
Returning condensed liquid enters the cryostat  through a vacuum insulated pipe. A small phase separator on the end of the pipe vents
the returning vapor to the cryostat vapor space. A submerged flex line carries the liquid to a sintered metal filter which then discharges into the bottom of the cryostat.
Passing liquid through such a cap has been known to improve the purity of the liquid argon as residual impurities present in the liquid can be adsorbed onto the cold sintered metal surface and therefore removed from the volume. \\
The fully outfitted {\sf ArgoNeuT} cryostat  can be seen in figure~\ref{cryo_setup}. The main cryostat and cryogenic system specifications can be found in table~\ref{cryo}. 

\begin{figure}[tb]
\begin{centering}
\begin{tabular}{c}
\includegraphics[height=3.6in]{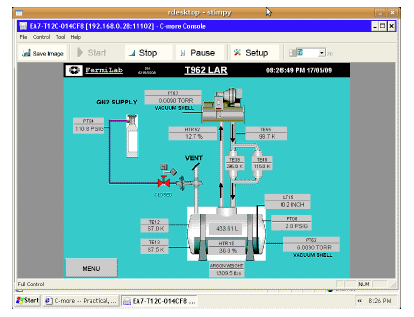} 
\end{tabular}
\caption{A screenshot of the remotely controllable cryo-system monitoring software applet of the {\sf ArgoNeuT} slow-control system, with display of the relevant thermodynamic parameters of the system during typical operations.}
\label{stimpy_screen}
\end{centering}
\end{figure}

The inner and outer vessels of the cryostat, the heat exchanger of the cryocooler, and the filters are instrumented with a number of temperature and pressure sensors. A set of heaters in the liquid at the bottom of the inner vessel and clamped to the heat exchanger are employed
 in order to to keep the load at exactly the entered set-point and also to promote recirculation. 
 Sensors and heaters are the main elements of the {\sf ArgoNeuT} slow-control layout. These are connectable and controllable locally, in the MINOS-ND hall, and remotely.\\
 The sensors, in combination with the heaters throughout the volume, make up a feedback loop. Slow-control software based on PID  (proportionalÐintegralÐderivative) algorithms controls this loop by actively adjusting the heater output such that a constant $\sim$140 mbar over-pressure (2.0~psig) is maintained in the gas pocket above the liquid inside the cryostat. If the pressure dips below 2.0~psig, heaters attached to the cryocooler kick on and the net cooling power of the device is reduced. If the pressure increases above this value, the heaters are automatically dialed down and the cooling allowed to increase. As there is an excess of cryocooling power above the heat load on the system, the heaters on the cryocooler are usually on and the system is consistently stable. A screenshot of the remotely controllable  {\sf ArgoNeuT} slow-control software at work  can be seen in figure~\ref{stimpy_screen}. \\
Relief lines are placed on every liquid argon containment volume in order to allow the argon to expand and escape the volume safely. All of the relief lines lead to a common vent pipe that is routed from the MINOS-ND hall up to the surface. The outer cryostat and surrounding containment vessel (the orange box in figure~\ref{detectorpic}) act as containers for any spilled liquid in the case of a major accident or failure. The containment vessel also contains oxygen deficiency sensors and fans.

\subsection{Liquid Argon Purification}
\label{purification}
The main limitation for a full collection of the free electron charge in ionization events is due to the presence, usually at trace level,
of electro-negative molecules in LAr. 
The electron attachment process to electro-negative impurities (like O$_2$, with $~e^-+~$O$_2\rightarrow ~$O$_2^-$) \cite{swan-biller}   
 is characterized by a high rate constant $k_e$ and the free electron charge exponentially decreases in time with a time constant $\tau_e$ 
 (the \emph{electron lifetime}) inversely proportional to the impurity concentration:
\begin{equation}
\frac{1}{\tau_e}~=~{k_e~[O_2]}
\label{eq:O2_conc}
\end{equation}
The total concentration of electro-negative impurities  in LAr
is normally reported in terms of the Oxygen equivalent [O$_2$] concentration and expressed in ppb units (parts-per-billion).\\
The value of the rate constant $k_e$ of the attachment process depends upon the drift field applied to the active LAr volume.
As a point of reference, $k_e$ assumes the value of $\simeq$ 3.1~ppb$^{-1}$ ms$^{-1}$ at electric field EF = 0.5 kV/cm \cite{bakale}. 
The electron drift velocity in a LAr at this field is about 1.6~mm/$\mu$s and, for example, when the impurity concentration is 0.7 ppb (and $\tau_e = 450~\mu$s) after a drift of 50 cm (drift time 310 $\mu$s) 
only 50\% of the free electron charge initially released by ionization survives capture (and can be collected).\\
Commercially available best grade liquid argon usually arrives with  an oxygen-equivalent concentration at the parts-per-million (ppm) level. The implementation of dedicated methods for removal of the residual impurity content in LAr is therefore necessary.\\
Filtration methods should match two conflicting requirements: high removal efficiency ($\gg~99.99$\%, in particular during the filling of the detector)
and fast operation time (considering the large amount of argon to process). This last requirement motivates the operation of  filtration in the liquid phase (exploiting a gas-to-liquid volume ratio $\simeq 800$ for argon from room to LAr temperature).\\
\begin{table}[h]
\centering
\begin{tabular}{|c|c|}
	\hline
Ar purification & Liquid phase (after GAr recondensation) \\ \hline
Ar Filtration Methods & O$_2$ catalyst (Cu-0226 S) + H$_2$O Molecular Sieve (Zeolite) \\ \hline
Filter Cartridges (custom) & Two $\times$  2 lt Volume \\ \hline
LAr purity and e-Lifetime & [O$_2$]$\approx 0.5$ ppb, $\tau_e$ $\approx $ 650 $\mu$s  \\ \hline
\hline
\end{tabular}
\caption{Main specifications of the {\sf ArgoNeuT} Ar purification system.}
\label{purif}
\end{table}

Several types of oxygen catalyst are commercially available and routinely employed in industrial processes 
where purified inert gas streams are used. 
Copper oxide or chromium oxide impregnated on high surface area chemically inert supports (like alumina or silica gel granules) are the most common materials.
After activation (or regeneration) through chemical reduction of the metal oxide into metal form,
filters for industrial applications made of non oxygen-permeable cartridges filled with catalyst pellets are used on-stream along the gas lines.\\
High efficiency oxygen removal is obtained by oxidation reaction with the metal to form the metal oxide 
and therefore permanently removed from the stream. 
The process is dependent on ambient temperature and pressure.\\
The removal efficiency is a function of the oxygen storage capacity of the filter that decreases with growing oxygen retention.
When the filter becomes saturated with oxygen and the efficiency drops below an acceptable level, full activity is restored by a simple
regeneration of the metal oxide back to metal form. This reduction process, also implemented for the initial activation, is carried
out by heating the filter with a flow of hot inert gas containing 2 - 4 \% of H$_2$ by volume.\\

Filters of this type have been custom made for experimental applications and used on-stream along liquid argon recirculation/reliquefaction lines in different LArTPC detectors since the early development of the liquid phase purification technique~\cite{ICAR-liqPhPur}.\\
The oxygen storage capacity of the reagent reduces considerably when operating at low temperatures, 
although this is partially compensated by the very low flow rate of the liquid through the cartridge (compared to gas-stream conditions) that increases the time available for the oxygen reaction to take place. \\
Adequate dimensioning of the filter size and prompt catalyst regeneration at efficiency breakthrough (when efficiency drops below a given threshold) have allowed electron lifetimes in the ms range ([O$_2$]$<$ 0.1 ppb) to be routinely achieved 
and maintained over extended run periods.\\

The main specifications of the Ar purification system can be seen in table~\ref{purif}. 
The {\sf ArgoNeuT} filters are made of activated-copper-coated alumina granules
inside of ConFlat\textsuperscript{\textregistered} flanged cylindrical stainless-steel cartridges. Liquid argon inlet/outlet flange-fitted lines at the cartridge ends are equipped with vacuum tight cryogenic valves for on-stream connection (disconnection) in the recirculation/reliquefaction closed loop system. Sinterized steel disks positioned at both ends of the internal volume serve to keep the granules inside the filter while letting liquid argon flow through. 
Each cartridge is 6.4 cm (2.5'') in diameter and 61 cm long, corresponding to about 2 L of volume (1.7 kg filter material, $\sim$10\% Cu). 
The active metal (Cu) surface available for O$_2$ reaction is very large (about 200 m$^2$/gr)
due to the porous structure of the alumina support. The alumina structure also provides H$_2$O adsorption capability by molecular trapping.\\
Presence of water  at ppm level in GAr (from material outgassing - plastic and G10 of cables, baffles and supports) in the cryostat neck at warm temperature has been observed to induce a drop in the electron lifetime when mixing into LAr.
To avoid this effect, one of the three filters made for the {\sf ArgoNeuT} recirculation system has been partially filled (33\%) with zeolite molecular sieve to increase water adsorption.

The cartridges are wrapped in about a 10 cm thick fiberglass insulation.
Two of them are installed in parallel just downstream of the cryocooler re-liquefaction vessel as active components of the purification loop pathway, with one open to LAr recirculation. When a filter reaches saturation, LAr circulation is switched through the other filter, while the saturated cartridge is replaced with the spare.  The exhausted filter material is then regenerated off-line and the cartridge is made ready to use again. \\
While in the recirculation system, the filters are low grade vacuum insulated to reduce the heat load. 
During steady operation, the liquid from boil-off argon re-condensation at the cryocooler is forced by gravity and pressure through the filter and returned purified back to the bottom of the cryostat. The recirculation system operates at a flow rate of about 1.6  lt/hr/lt (LAr volume per unit of time per volume of filter) providing a full volume exchange every 7-8~days ($\sim$550~liters of LAr content in the cryostat). 
The estimated oxygen capacity for liquid argon filtering is 0.5 g O$_2$ per kg filter media.  Filter saturation and replacement during the {\sf ArgoNeuT} run period occurred every three weeks on average.\\

The series of two O$_2$ filters of the same type as those in the recirculation system (one with 33\% molecular sieve content for water adsorption) has been included in the filling line utilized during detector filling operations for a first and substantial reduction of the O$_2$ content in the delivered commercial argon. \\

A full description of the {\sf ArgoNeuT} filter technique using a non-proprietary and regenerable design can be found at ~\cite{curioni}.

\subsection{Time Projection Chamber}
\label{sec:TPC}
Charged particles crossing a volume of liquid argon produce free electron tracks by ionization.
The LArTPC technology allows
for three-dimensional image reconstruction and
calorimetric  measurement of the ionizing events. \\
A  uniform electric field 
is applied to the medium and the ionization electron tracks are
projected onto the anode along the electric field lines. 
 The read-out of the electron track image is obtained by configuring the anode as a system
of parallel wire-planes (number of planes $\ge 2$), biased at specific potentials to enhance "transparency" 
of the successive wire plane to drifting electrons.
With this configuration, each segment of a track induces a pulse signal (``hit") on one wire in each plane (in normal cases). 
The coordinate of the wire in the plane provides the hit position, so that multiple and independent localizations of the track segment  can be accomplished  (``non-destructive" read-out)~\cite{gatti}. 
Timing of the pulse information, combined with the 
drift velocity information, determines the drift-coordinate of the hit, thus providing full three dimensional (3D) image reconstruction capability.\\
In liquid argon, no charge multiplication occurs. The signal pulse height is  therefore 
proportional to the amount of ionization charge in the track segment. A precise measurement of the deposited energy along the track can thus be extracted by summing the charge over the entire track length in LAr. \\
\begin{figure}[h]
\begin{centering}
\includegraphics[height=4.0in]{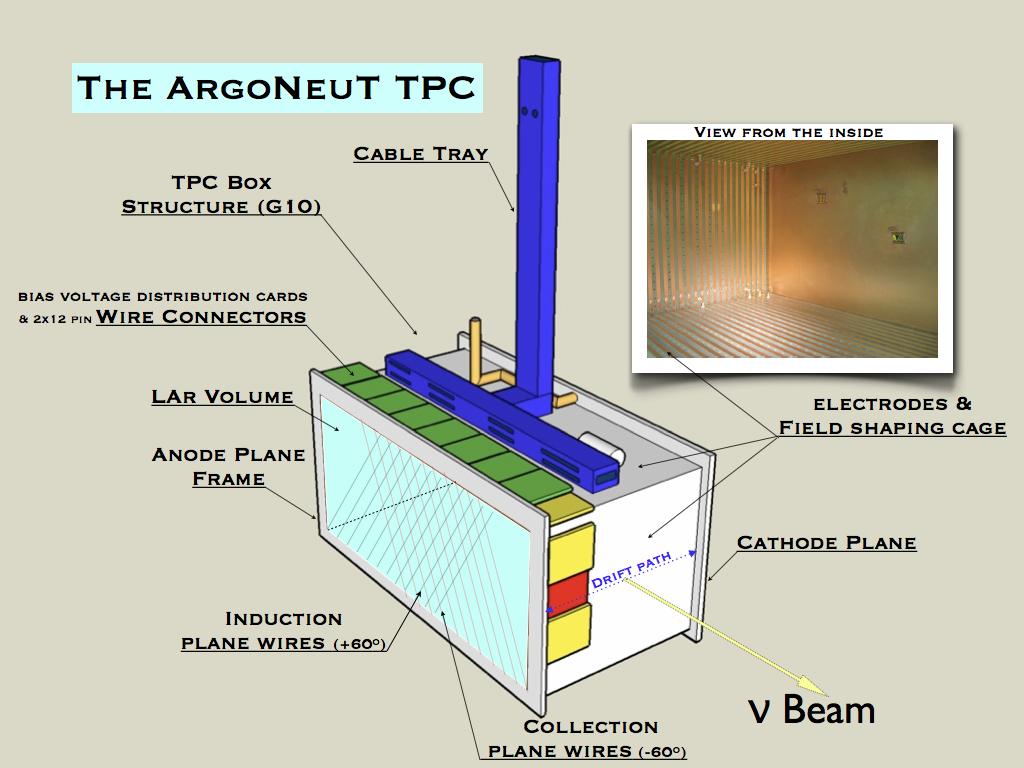}
\caption{Pictorial view of the {\sf ArgoNeuT} LArTPC mechanics. Details of the anodic structure with the ($\pm 60^o$) inclined wire-planes are indicated.
In the insert, a picture of the inside of the LArTPC volume showing the cathodic plane and the copper strips of the field shaping cage.}
\label{TPC_details}
\end{centering}
\end{figure}

A pictorial sketch of the {\sf ArgoNeuT} LArTPC is shown in figure~\ref{TPC_details}. The main TPC specifications can be found in table~\ref{TPC}. 
The active volume is  $40~h\times 47~w\times 90~l~$cm$^{3}$, corresponding to about $\sim$170~liters of LAr. This volume is delimited by a rectangular box structure sitting inside of the cryostat. The structure itself, frame and side panels, is composed of G10, a composite material comprised of woven fiberglass with an epoxy resin binder.  The chamber is oriented so that the longest dimension is approximately parallel to the beam. Two opposite sides of the box (40$\times$90~cm$^{2}$) are instrumented as cathode and anode planes of the TPC, with the drift direction horizontal with respect to the ground and perpendicular to the beam. 
Thus, the maximum drift length inside the TPC volume is $\ell_d=47$ cm, from the negatively charged cathode to the first, innermost 
wire-plane of the anode system.\\
The cathode is a G10 plain sheet with copper metallization on the inner surface. 
The anode of the {\sf ArgoNeuT} TPC is made of a series of three parallel wire-planes. The three planes are equally spaced with interplane gaps of $\ell_g=4$ mm. \\
Wires are made of Beryllium-Copper Alloy \#25,
with a diameter of 152 $\mu$m (0.006''), strung at a nominal tension force of 9.8 N by a wire-winding machine. 
Anchorage is provided by soldering the wire ends onto copper pads arranged on the rectangular anode wireplane frame of the TPC box. 
A series of G10 spreader bars
are externally added to the frame to reinforce the mechanical structure against wires tensioning and thermal stresses. 
\begin{table}[h]
\centering
\begin{tabular}{|c|c|}
	\hline
TPC dimensions  &  $40~h\times 47~w\times 90~l~$cm$^{3}$  \\ \hline
TPC (active) volume  &170 liters  \\ \hline
Max. Drift Length (TPC width) &  $\ell_d=470$ mm \\ \hline
\# of wire-planes & 3 (2 instrumented - I, C)  \\ \hline 
Interplane gaps width &$\ell_g=4$ mm \\ \hline 
Wire pitch (normal to wire direction) & $\delta s=4$~mm (all planes)  \\ \hline
Wire Type & Be-Cu Alloy \#25, diam. 152 $\mu$m  \\ \hline
\# of wires (total) & 705 \\ 
~~~~~~Shield plane (S) & 225  (non-instrumented)\\ 
~~~~~~Induction plane (I) &  240  (instrumented -  w-index: $n_w^I$) \\ 
~~~~~~Collection plane (C) & 240  (instrumented - w-index: $n_w^C$)\\ \hline
Wire Orientation (w.r.t. horizontal) & $90^o,~+60^o,~-60^o$ (S,~I,~C) \\ \hline 
Non-destructive Configuration &  EF nominal (Transparency Ratio) \\ 
~~~~~~ Drift volume  & $E_d=500$ V/cm  \\
~~~~~~ S-I gap & $E_{g1}$=700~V/cm  ($r_T=1.4$) \\
~~~~~~ I-C gap & $E_{g2}$=900~V/cm  ($r_T=1.3$) \\ \hline
Drift Velocity (at nominal field) & 1.59 mm/$\mu$s \\ \hline
Max. Drift Time (at nominal field)  & $t_d=295~\mu$s \\ \hline
\hline
\end{tabular}
\caption{{\sf ArgoNeuT} LArTPC nominal specifications and features. 
Operational values of the fields configuration and related parameters adopted during the physics run differ slightly from
these nominal values due to further optimization in experimental conditions.}
\label{TPC}
\end{table}
The tension force on the wires is increased by an additional $\sim$ 5 N via the spreader bars after wire plane assembly in the TPC frame.
The wire spacing (pitch) is $\delta s=4$~mm in all planes.\\
The first plane contains 225 parallel wires, vertically oriented with respect to the ground and perpendicular to the beam axis ($\mathrm{+90^{\circ}}$). This plane is not instrumented for readout purposes but serves to shape the electric field near the wire-plane and to shield the outer, instrumented planes against induction signals from the ionization charges while they are drifting through the TPC volume. This first plane is thus indicated as  the ``Shield plane".
Wires are all equal in length (40.0 cm) and stretched between the horizontal edges (top and bottom) of the frame. \\
The second plane, indicated as the ``Induction plane",
consists of 240 wires all oriented at $\mathrm{+60^{\circ}}$ relative to the beam axis. 
Electrons induce signals on the wires of this plane only after crossing the Shield plane and moving toward, across and away from the Induction plane, forming a bipolar shaped current pulse. \\
The third (last) plane is made up of 240 wires oriented at $\mathrm{-60^{\circ}}$ relative to the beam axis. Electrons are collected onto the wires of this plane at the end of their drift (unipolar current pulse), and the plane is thus indicated as the ``Collection plane". \\
Inclined wires ($\pm \mathrm{60^{\circ}}$) of the Induction and Collection planes are of varied length. 
The majority of the wires (144 in each plane), with equal length of 46.2 cm, are stretched between the horizontal edges of the frame, while wires of decreasing length are used in the two triangular shaped portions, 
between one vertical and one horizontal side, at the corners of the wire plane area (48+48 wires in each plane). 
The shortest wire length is 3.7 cm, and only the extreme corners (top-left and bottom-right for the Induction, bottom-left and top-right for the Collection) of the inclined wire planes are left uninstrumented. 
The orientation of the wires in the Induction and Collection planes can be seen in figure~\ref{TPC_details}. 

The reference value of the electric field throughout the drift region of the TPC is $E_d$=500~V/cm.
This can be obtained by biasing the TPC cathode at negative high voltage, with the
HV generated by an external low-ripple power supply connected to the inner cathode through a vacuum tight HV feed-through (HV-FT). The HV-FT is custom made by Fermilab with
a conductor rod ($\o$=16 mm) forced tightly into the hole of a high-density polyethylene (HDPE) tube. 
After cooling with liquid nitrogen (LN$_2$), the tube is forced into a stainless steel tube flanged at one side. Vacuum tightness is obtained by thermal expansion of the HDPE returning to room temperature. The tube with the HDPE insulator and the conductor rod is about 1.2 m in length. From a dedicated ConFlat\textsuperscript{\textregistered} flanged port on the top flange of the cryostat chimney, the HV feed-through dips down into the LAr volume. The conductor lead of the HV-FT is then connected with the cathode of the TPC by a flexible conductor rod bolted at both ends. \\

A ``non-destructive" read-out configuration of the anode system of wire planes can be established by biasing the planes at suitable potentials, such that wire-plane ``transparency" to drifting electron charges across Shield and Induction planes is maximized. \\
Transparency is a function of the wire geometry (diameter and pitch) and of the electric fields in the interplane gaps. 
Transparency enhanced above geometrical value (96 \%) is obtained for $E_{g2}\ge r_{T2}~E_{g1}$ and $E_{g1}\ge r_{T1}~E_{d}$, where $E_{g1,2}$ are the field values in the first and second gap between the Shield and Induction and between the Induction and Collection planes respectively, with $1.1\le r_{Ti}\le 1.5$ the range of the field scaling factor usually required to obtain good transparency.\\
{\sf ArgoNeuT} reference fields values in the two gaps are $E_{g1}$=700~V/cm  ($r_{T1}=1.4$) and $E_{g2}$=900~V/cm  ($r_{T2}=1.3$).
These fields can be established with low bias voltages applied to the wire planes of the TPC by an external DC power supply (negative low voltage
for Shield and Induction planes and positive low voltage for the Collection plane).\\
Actual values of the fields configuration adopted during the physics run after electronic noise minimization are close to the reference values given above and will be reported in Sec.\ref{sec:commissioning}.\\
\begin{figure}[tb]
\begin{centering}
\begin{tabular}{c}
\includegraphics[height=3.in]{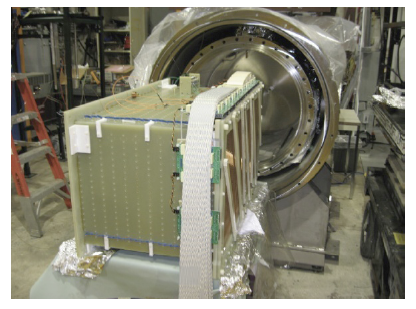} 
\end{tabular}
\vspace{0cm}
\caption{The fully instrumented TPC being inserted into the {\sf ArgoNeuT} cryostat opened through the removed front-end cap. 
The cryostat inner and outer vessels and the vacuum jacket in between are visible.}
\label{tpc_taup}
\end{centering}
\end{figure}
The electric field is made uniform over the entire TPC drift volume by means of a field shaping system of electrodes placed onto
 the boundary surface surrounding the volume between the cathode and anode planes (see insert of figure~\ref{TPC_details}). 
For this purpose, the four G10 side panels delimiting the TPC box have been manufactured by PCB technique with copper strips 1~cm wide spaced at 1~cm intervals, forming 23 rectangular rings all the way up the TPC. Copper tabs soldered in the four corners of the rings provide a solid electrical connection of the copper strips.
The rings are set at a potential linearly decreasing from the cathode to the Shield plane. This shapes the field uniformly inside and near the edges of the TPC volume, and hence ionization electrons may move at constant drift velocity. The biasing potentials of the rings 
are obtained through a chain of resistive voltage dividers. 
Four parallel  resistors, 100 M$\Omega$ each, are used between the cathode and the first strip ring electrode and between every pair of adjacent rigs. A last set of four resistors connect the last ring to ground. The resistor chain is thus made of 24 steps with R=25 M$\Omega$ per step,
corresponding to a total chain resistance of 600 M$\Omega$ from cathode to ground. 

The inside of the TPC can be seen in figure~\ref{TPC_details} where the cathode plane and a portion of the 
field-shaping system are shown. The entire TPC is visible in figure~\ref{tpc_taup} with a picture taken just before insertion inside the cryostat.

\section{Readout Electronics}
\label{ROelectronics}
\begin{figure}[h]
\begin{centering}
\begin{tabular}{c c}
\hspace{-0.5cm}
\includegraphics[height=2.9in]{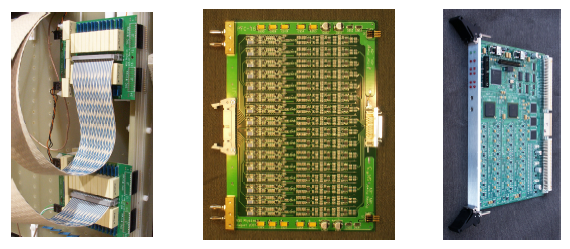} 
\end{tabular}
\caption{[Left] A set of Bias Voltage Distribution Cards (BVDC) mounted onto the TPC frame and connected with the internal readout cables routing the TPC wire signals toward the electrical feedthrough flange. [Middle] The Preamplifier and Filter Card (PFC). [Right] The ADF-2 digitizer card.}
\label{electronics_preamp_other}
\end{centering}
\end{figure}

The readout electronics implemented for the {\sf ArgoNeuT} physics run is structured
as a multi-channel waveform recorder that continuously records charge information collected 
by each sense wire during the drift of ionization electrons inside the TPC.\\
The readout chain for the wire signals is composed of a series of three electronics cards and boards: 
 (1) the bias voltage distribution (and decoupling) card (BVDC), 
 (2) the preamplifier and filter card (PFC-16) and 
 (3) the ADC, circular memory buffer and VME readout digitizer module (ADF-2).
 The main read-out electronics specifications can be seen in table~\ref{RO-elec}.
 figure~\ref{electronics_preamp_other} displays a picture of each type of active component of the read-out chain. \\
 Connecting elements are the internal readout cables, the readout signal feedthrough card, the external readout cables and the pleated foil cables. 
\begin{table}[h]
\centering
\begin{tabular}{|c|c|}
	\hline
Bias Voltage Distribution Card  (BVDC) & 20 units \\ 
~~~ \# channels per BVDC & 24 \\
~~~  DC Blocking  Capacitor & 10 nF (1600 Volt)\\  \hline
Preamplifier and Filter Card (PFC-16) & 30 units \\ 
~~~ \# channels per PFC & 16 \\
~~~ PreAmp stage - FET Voltage Gain & 0.5 mV/fC\\
~~~ Shaping \& Filter Stage  &  {\it ``Narrow Gaussian"} \\
~~~ 1-pole High Pass  & $\tau_{diff} \simeq 3~\mu$s; ~~$f^c_{HP}=55$ kHz \\
~~~ 2-pole Low Pass $\times$ 2  &  $\tau_{int} \simeq 1~\mu$s;~~$f^c_{LP}=160 $ kHz\\  \hline
Digitizer Module (ADF-2) & 15 units \\ 
~~~ \# channels per ADF-2 & 32 \\
~~~ ADC range & 10 bit \\
~~~ ADC Gain & 0.1881 ADC/mV \\
~~~ Sampling Time (FPGA) & $\delta t$=198 ns (5.05 MHz) \\
~~~ Time ticks per record & $n_t$= 1,..., 2048 \\
~~~ Record Length & 405.5 $\mu$s  \\ 
~~~ \# of Pre-Sampling ticks & 60 (11.88 $\mu$s) \\ \hline
Electronics Charge Sensitivity & 7.49 ADC/fC  \\ 
Tot. Capacitance (Det. and Cables) & 230 pF \\
Response to mip (Coll. wires)  & $S/N \ge 15$ \\ \hline
\hline
\end{tabular}
\caption{Some relevant {\sf ArgoNeuT} read-out electronics specifications.}
\label{RO-elec}
\end{table}

Bias voltage from a heavily noise-filtered external DC power supply is daisy-chained to each BVDC  (20 units), passed through an additional noise filter and distributed to each sense-wire of the TPC through a 100~M$\Omega$ isolation and current limiting resistor. To limit ground-loop currents, the ground bus of the daisy-chain is isolated from the signal ground on the BVDC by 100~$\Omega$ resistors. Between each TPC wire and the readout cable connector is a DC blocking (10 nF, 1600 Volt) capacitor that decouples the DC bias voltage from the AC wire signal. The resistors and capacitors are mounted on the BVDC, which are placed directly on the TPC mechanical structure (see figure~\ref{tpc_taup}) inside the cryostat. Each BVDC services 24~TPC wires and connects to the TPC wires with two 12-pin connectors. The BVDC are submerged in the liquid argon when the cryostat is filled. BVDC prototypes were tested in liquid argon at the Fermilab Materials Test Stand \cite{FNAL_MTS}, with no observable effect on electron lifetime. \\

Twenty internal readout flat cables (halogen free polyolefin insulation material, tinned stranded copper conductor) each with 24 signal/ground twisted wire-pairs are packed inside a dedicated cable-tray, and carry signals from the BVDC up the chimney of the cryostat to the inner side 
of the signal feedthrough. \\
The internal readout cables are about 2.7 m long, with a characteristic capacitance of 48 pF/m, are in argon gas for most of their length in the upper part of the cryostat to the feedthrough. \\

\begin{figure}[h]
\begin{centering}
\begin{tabular}{c c}
\includegraphics[height=2.2in]{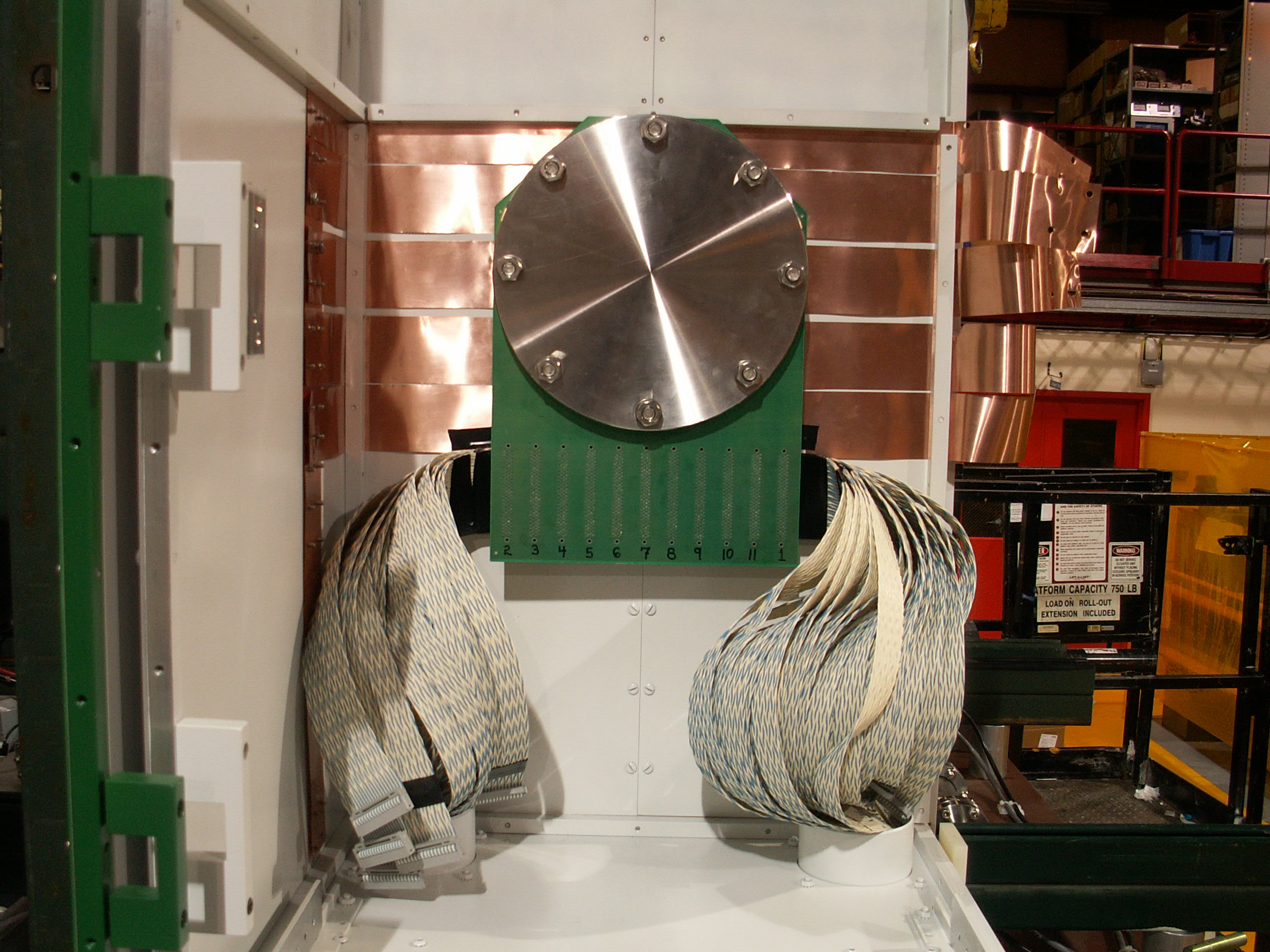} 
\hspace{0.5cm}
\includegraphics[height=2.2in]{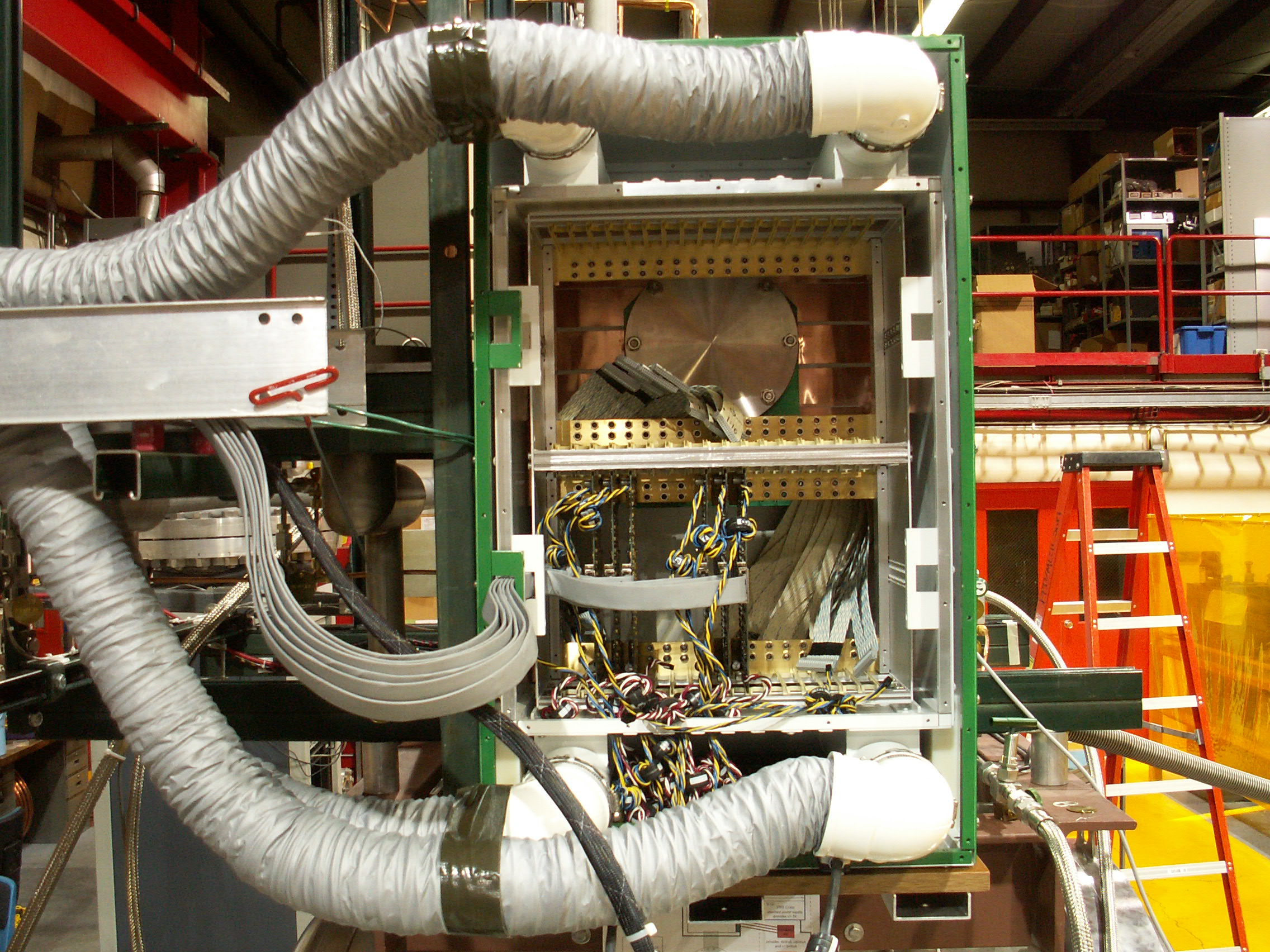} 
\end{tabular}
\caption{View of the readout signal feedthrough flange inside the RF shielded cage (partially) mounted on the outside of the cryostat. The external readout cables for the connection from the feedthrough PC board to the Preamplifier and Filter Cards are also visible in the picture. The box with 
PFC in their crate and the pleated foil cables to the digitizer board are shown on the [Right].}
\label{fig:feedthrough_installed}
\end{centering}
\end{figure}

A readout signal feedthrough bridges the gap between the vacuum tight cryostat and the outside world. It is custom made with board-mount plug connectors (11 connectors of 100 pins each) inside and outside of the cryostat. The internal connectors are mounted on a part of the feedthrough printed circuit board that is captured between the flange of a dedicated port on the cryostat chimney and a blanking flange plate, visible in figure~\ref{fig:feedthrough_installed}. 
This creates a vacuum tight seal around the internal connectors.
The external connectors are located on an accessible portion of the feedthrough board that extends beyond the flange, with the pins of corresponding in-out connectors connected by traces within the feedthrough board.\\

 After exiting the feedthrough, signals are transported to the 16 channel preamplifier and filter card (PFC-16) via short (about 30 cm) cables of the same type of polyolefin insulated ribbon cable used inside the cryostat.\\
  
The PFC-16 card has a preamplifier stage and a shaping/filter stage.
The preamplifier stage (dual FET integrating charge to voltage preamplifier) is similar to the FET stage of the D-Zero Run II hybrid preamp \cite{D0-electronics}.  It is setup with a 2.0~pF effective feedback capacitor (i.e. 0.5 mV output per fC input) and a 100 M$\Omega$ feedback resistor (time constant of 200~$\mu$s) to prevent saturation. 
This is followed by a buffer stage converting the preamplifier output to a differential analog signal with a voltage gain of 8.25.\\
At the output of the buffer stage a narrow Gaussian shaping filter is implemented as one stage of differentiation, followed by two stages of integration in the filter section. The differentiation stage provides a high-pass filter with a frequency cutoff of $\sim$55 kHz ($\tau_{diff}\simeq 3~\mu$s). This sets the fall time of the filter output for unipolar signals (e.g. from the collection plane wires).
  The filter section of the PFC-16 has two differential stages of filtering. Both are two-pole low pass filters with a frequency cutoff of 160 kHz ($\tau_{int}\simeq 1~\mu$s) and a voltage gain of 6.0. The peaking time for unipolar signals is thus $\sim 2 \tau_{int}$.\\
  Pulse-shaping filters reduce the total noise but also decrease the peak signal amplitude. 
  With an integration-to-differentiation time constant ratio $\simeq 1/3$,  a $\sim$45\% loss in peak signal amplitude is
  estimated at the shaper output. 
  
The outside of the signal feedthrough flange, the External Readout Cables and the PFC-16 cards (30 units) are enclosed in a double RF shielded cage mounted on the outside of the cryostat. The cage features a remote duct cooling system for the preamplifier. \\
 
After amplification and filtering, the analog signal is routed along 7.5 m of pleated foil cables and digitized by a set of 32-channel ADF-2 modules. These modules were borrowed from unmodifiable D-Zero Run II trigger system spares. \\

Each ADF-2 module features a 10~bit analog-to-digital converter (0:1023 counts output). The ADF-2 channels have a capacitive coupling on inputs and six different sensitivities, which varied from 2.072 to 5.317 mV/bit. To compensate for these variations, the back-termination resistor on each PFC-16 channel was chosen to equalize the overall gain to the ADF-2 channel with the lowest sensitivity, resulting in a channel sensitivity of 1 ADC count per 5.317 mV at its input (0.1881 ADC/mV gain). An unintended consequence of this procedure was to create six slightly different signal shapes, with the variation  primarily appearing in a slow restoration of the baseline. Channel-by-channel corrections for these variations are incorporated in the digital signal processing performed as the first stage of the analysis software, 
as reported in Sec.\ref{raw_wfm}.\\
Analog signals from the PFC are "back terminated" and parallel terminated on the ADF-2 cards, thus the ADF-2 card sees only about 1/2 of the unterminated output signal from the last stage on the PFC-16 cards. \\
The ADF-2 modules are capable of greater than a 100 MHz sampling frequency but, for ArgoNeuT, the ADF-2 sample the preamplifier output every $\delta t$=198~ns  and record the digital information in a circular buffer for 2048 samples (or time ticks, $n_t$) implemented in a field-programmable gate array (FPGA). The recorded time window (405~$\mu$s) encompasses the maximum drift time ($\sim$300~$\mu$s at the nominal electron velocity in LAr) from the cathode to the wire planes. The extra time allows the events to be pre-sampled (60 time ticks) and post-sampled, providing a tool for actual baseline and noise evaluation on event-by-event basis. \\

This PFC-16 and ADF-2 read-out chain services all the channels of the TPC (Induction and Collection wires).
The corresponding ($2\times 240$) digital signal waveforms ($V(t)\equiv \left\{ V_{n_t}\right\}_{n_t=1,...,2048}$ in ADC counts) are then read out by a computer and saved in binary format, one file for each event/trigger.\\
 The baseline of each channel is adjusted to 400 counts at DAQ initialization. This gives a similar dynamic range for positive or bipolar incoming signals.\\

Based on the nominal characteristics of the electronics components and taking into account signal 
losses due to pulse shaping and to unmatched terminations, the overall sensitivity of the read-out chain 
is estimated to be 7.6 ADC counts per 1 fC at its input  (narrow current pulse). \\
From bench-tests of the electronics response, performed before electronics installation in the {\sf ArgoNeuT} experimental layout and thus 
without a definitive model for the actual ionization pulse from the wires, the sensitivity was found to be 
in reasonable agreement (slightly higher) than the expected value quoted above.\\
At the end of the physics run and through a dedicated off-line data analysis, a detector ``self-calibration" method was developed and applied \cite{ArgoNeuT-mu-paper}.  Assuming the theoretically determined most probable value of the energy deposited by crossing muons, the 
sensitivity of the read-out chain (the detector calibration factor $f_{self-cal}$ in \cite{ArgoNeuT-mu-paper}) is found to be 7.49$\pm$0.09~ADC/fC,
in good agreement with expectations.\\
    
The electronics noise for a total input capacitance of $\simeq 230$ pF,  which includes contributions from detector wire, internal/external cables and feedthrough, is expected to be in the range of 1.5 ADC counts per channel. A minimum ionizing particle releases about 3.5 fC/wire, corresponding to 
26 ADC/wire in the Collection plane and about 14 ADC/wire in the Induction plane, and therefore a minimum signal to noise ratio (S/N) larger than ten is expected.

A thorough description of the {\sf ArgoNeuT} electronics readout configuration can be found in Reference~\cite{edmunds}.

\subsection{External Scintillator System}
\label{sec:extscint}
\begin{figure}[h]
   \centering
    \includegraphics[width=2.5in]{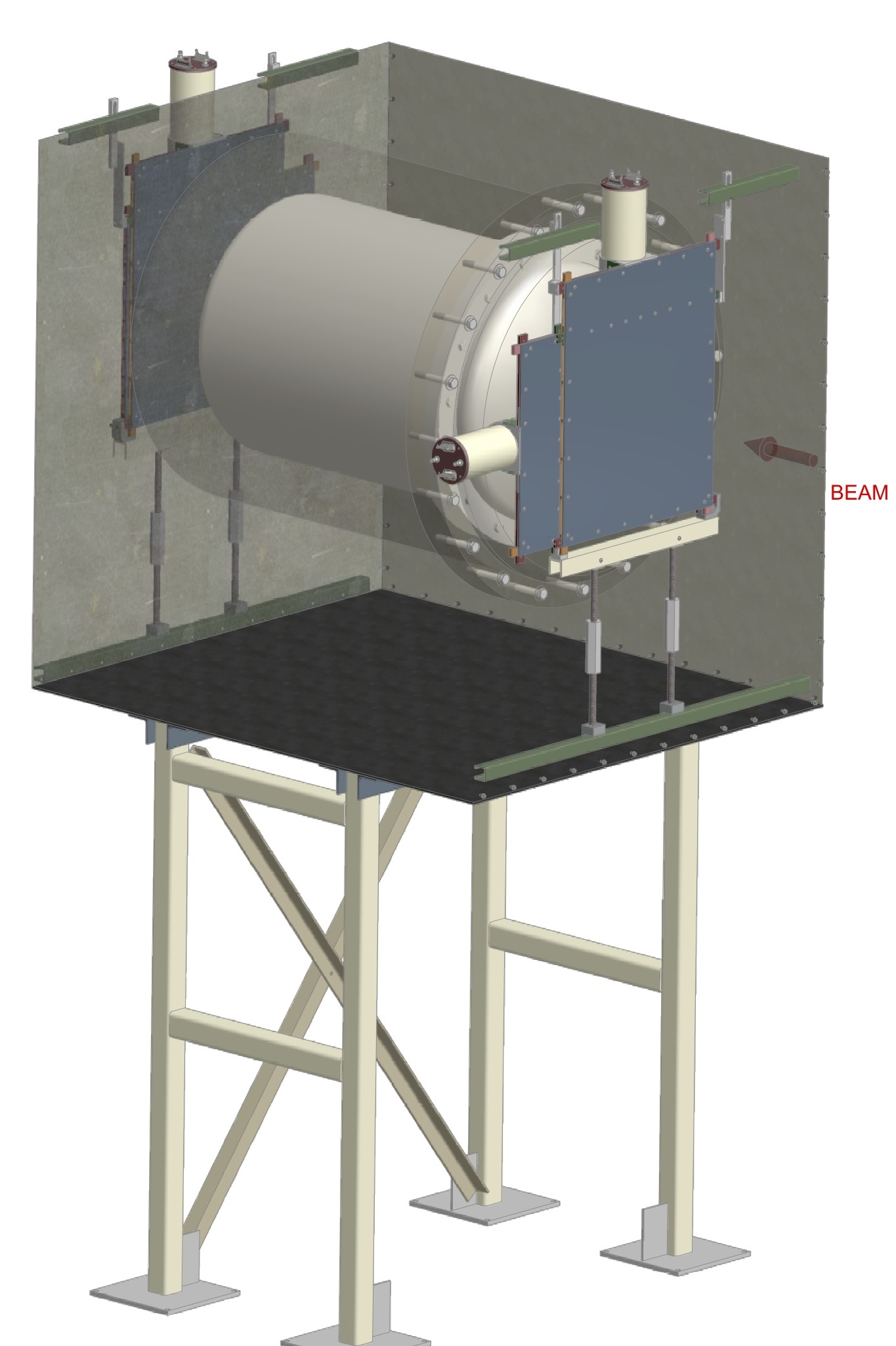}
    \includegraphics[width=3.4in]{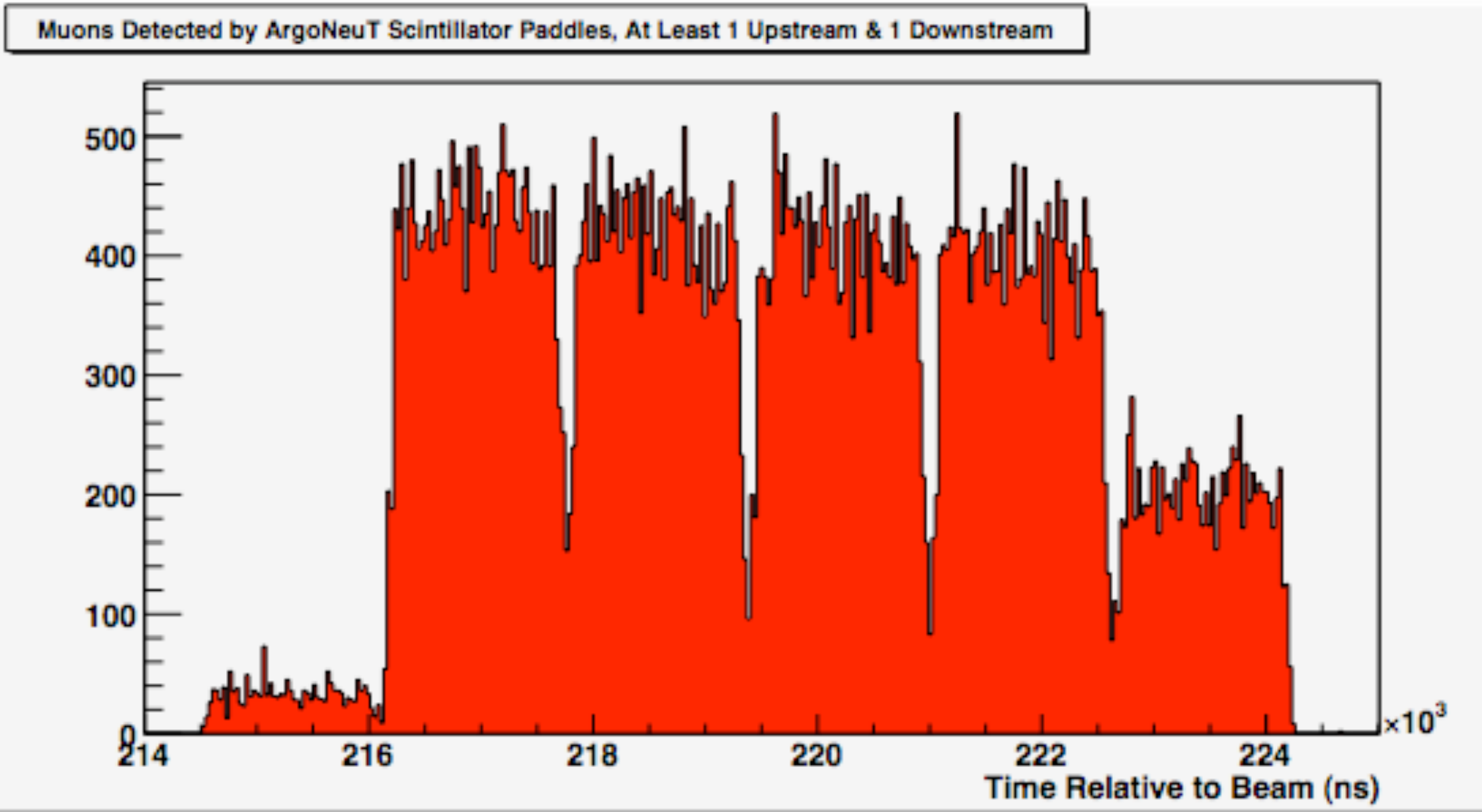} 
   \caption{[Left]  Scintillator paddles are arranged with two upstream and two downstream of the {\sf ArgoNeuT} cryostat. [Right] Timing for spills that triggered at least one upstream and one downstream {\sf ArgoNeuT} scintillator paddle. }
   \label{fig:location}
\end{figure}

The {\sf ArgoNeuT} experimental layout is complemented by  a system of four planes of scintillator (``paddles") to augment the trigger system.  
\begin{figure}[h]
   \centering
   \includegraphics[width=2.2in]{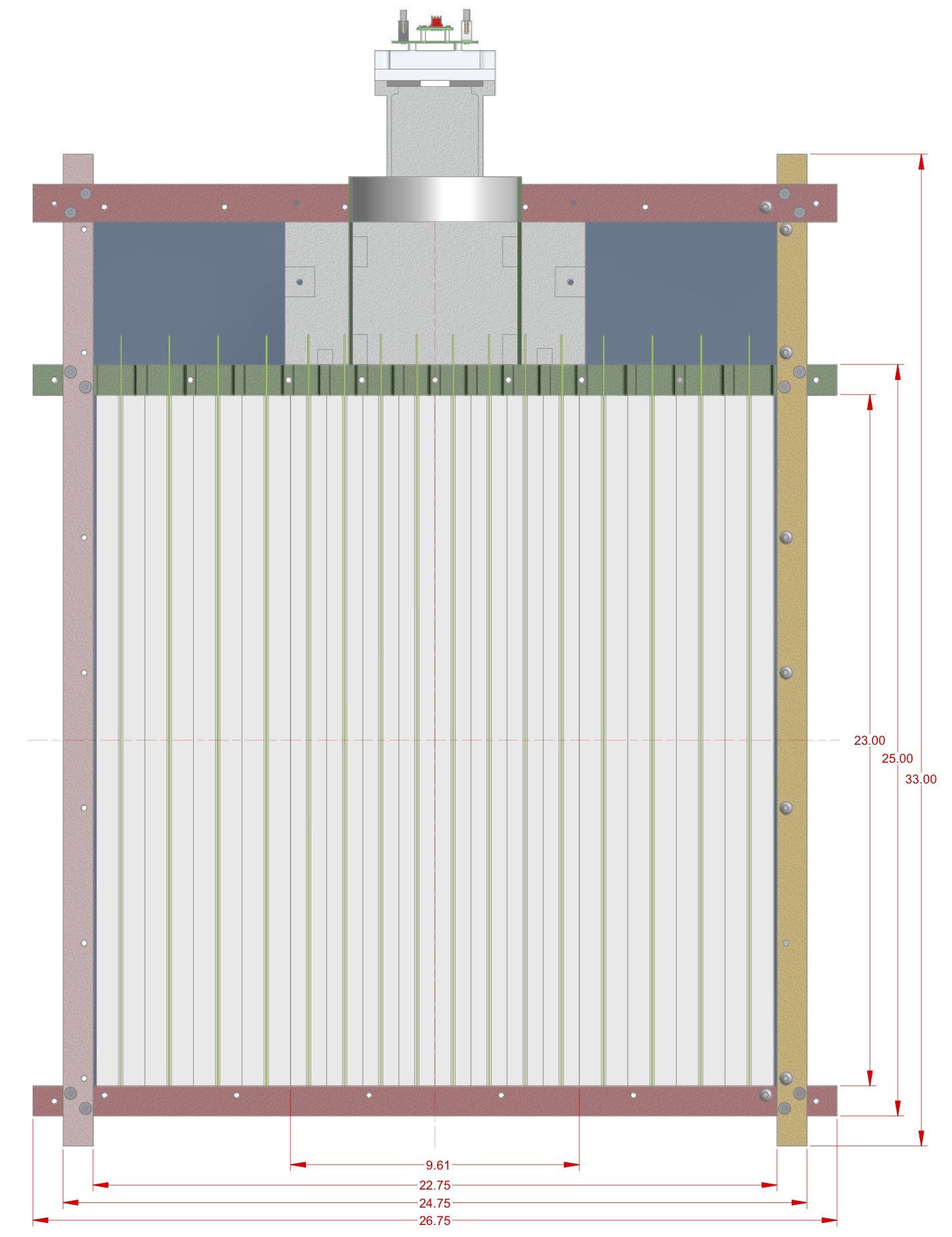} 
    \includegraphics[width=2.0in]{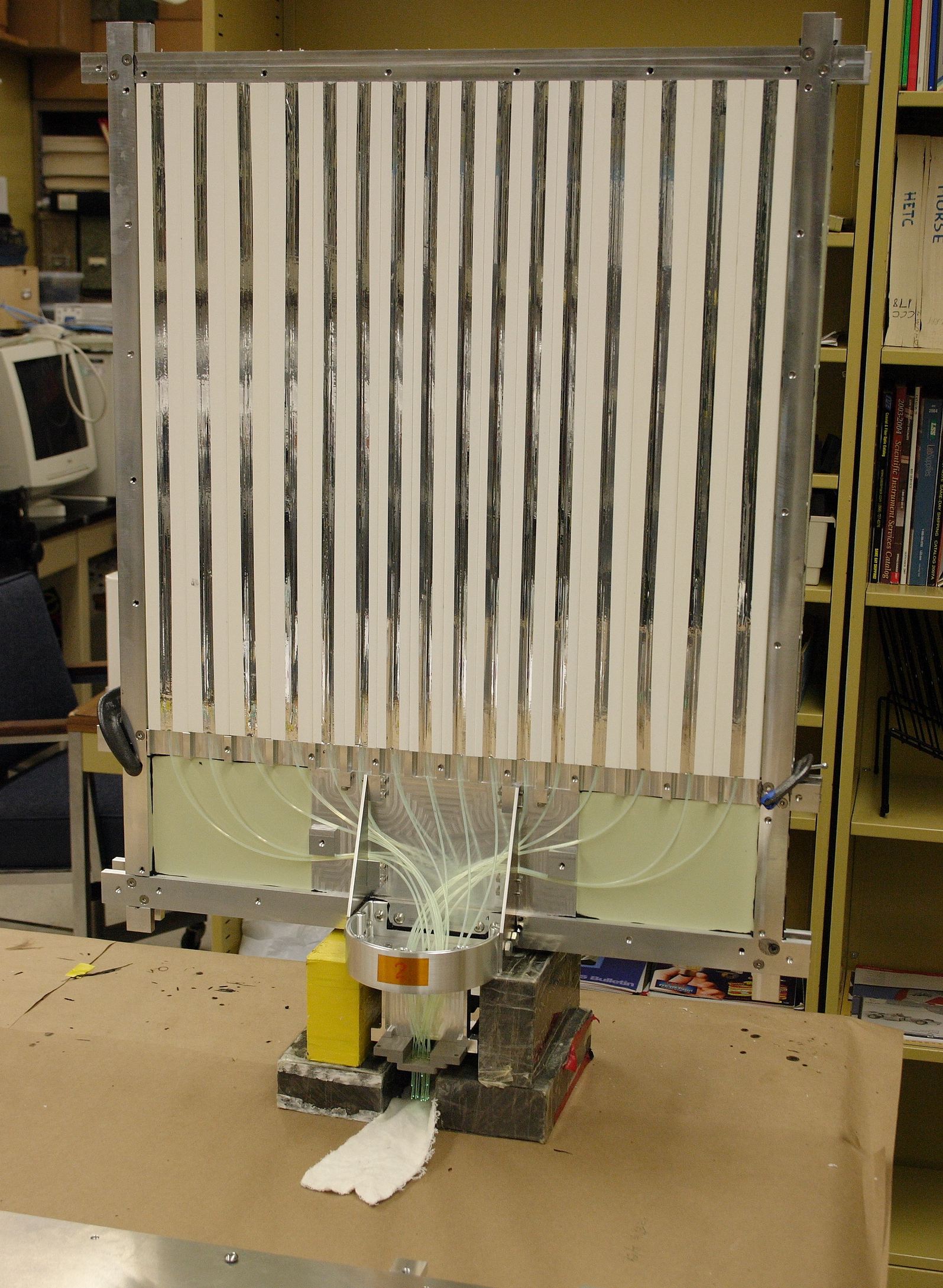}
   \caption{[Left] Schematic drawing of {\sf ArgoNeuT} scintillator paddle design, and [Right] a paddle under construction.}
   \label{fig:paddles}
\end{figure}
Each paddle consists of 16 scintillator bars with wavelength-shifting fibers glued onto the bars.  The scintillator bars are 58.5 cm in length, with the innermost eight scintillator bars 3 cm in width and the outermost eight scintillator bars 4 cm in width.  The fibers from all 16 scintillator strips of the paddle are routed to a multianode photomultiplier tube, as shown in figure~\ref{fig:paddles}.  The photomultiplier tubes (one per paddle) are operated at a slightly different (negative) voltage bias adjusted to achieve equal single count rate ($\approx$30 Hz).  A pair of paddles with scintillator strips oriented at 90$^{\circ}$ with respect to one another are located on both the upstream and the downstream end of the cryostat, as shown in figure~\ref{fig:location} [Left].  The paddles are approximately centered on the TPC area.
The signals from the scintillator paddles are passed to NIM discriminators, which in turn pass their output signals to a Time-To-Digital Converter board in a CAMAC crate.  Arrival times of pulses above threshold are recorded relative to the arrival of the signal from the Fermilab accelerator complex, which indicates delivery of protons to the NuMI target.  Figure~\ref{fig:location} [Right] shows the time in which  at least one upstream and one downstream {\sf ArgoNeuT} scintillator paddle recorded signals above threshold with respect to NuMI spills. The characteristic structure of the NuMI beam \cite{NuMI-beam}, which is delivered in either 5 or 6 batches per spill (in 5-batch mode either the first or last batch is not delivered to NuMI) over a $\Delta t_{spill}$=9.7$~\mu$s window, is clearly evident.  The delay of $\approx$215$~\mu$s between the accelerator signal and the beam arrival time as well as the spill duration are consistent with expectation.      

The scintillator paddles are not used to trigger on events, but rather are intended to reduce the uncertainty on the absolute time  of interactions ($t_0$) detected by the TPC.  Since there is no internal light collection in {\sf ArgoNeuT}, $t_0$ cannot be determined to a resolution smaller than the $\Delta t_{spill}$ beam window without incorporating some external constraint.  By combining the TPC information with the scintillator paddle information, the uncertainty on the $t_0$ can be substantially reduced leading to improved resolution on the drift-coordinate of particle tracks, which in turn improves the ability of {\sf ArgoNeuT} to match tracks to the MINOS-ND.  Studies are underway to use the scintillator information for this purpose.

\subsection{The MINOS Near Detector as a back muon catcher}
\label{sec:minos-nd}
The MINOS Near Detector (MINOS-ND), located just downstream of {\sf ArgoNeuT}, is an essential component in the {\sf ArgoNeuT} experimental layout, as it is utilized as a back muon catcher in the neutrino event reconstruction. \\
\begin{figure}[h]
\begin{centering}
\begin{tabular}{c}
\hspace{-1.3cm}
\includegraphics[height=4.0in]{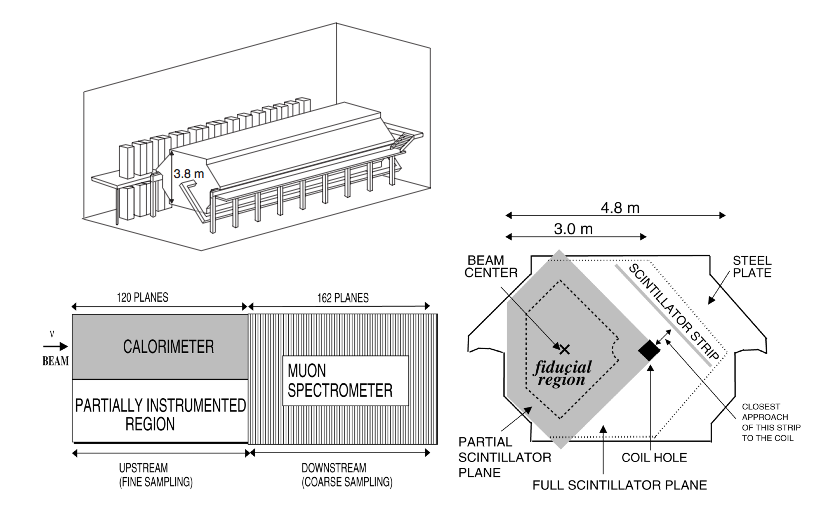} 
\end{tabular}
\caption{[Top] The MINOS-Near Detector at Fermilab.
[Left] The regions of the MINOS-ND. The ``partially instrumented" region has a full coverage plane every fifth plane. [Right] The MINOS-ND plane configurations. The beam and fiducial volume are centered around the middle of the partially instrumented planes, left of the coil. Each plane provides a two dimensional view and adjacent planes are combined to form a three dimensional image of the event. The plots are taken from Reference~\cite{MINOS-ND-tech}.}
\label{minosregions}
\end{centering}
\end{figure}

MINOS-ND is a magnetized steel/scintillator tracking/sampling calorimeter. 
A detailed description is available in ref. \cite{MINOS-ND-tech} by the MINOS Collaboration and the main features are summarized in table~\ref{tab:MINOS-ND}.
\begin{table}[h]
\centering
\begin{tabular}{|c|c|}
	\hline
MINOS-ND & magnetized tracking/sampling calorimeter \\  \hline
Shape/Dimensions & Squashed Octagon - $w$=6.2 m, $h$=3.8 m, $l$=12.8 m \\
Mass & 980 t  \\ 
Magnetic Field & 1.3 T (toroidal) \\ \hline
Detector configuration & steel plate/plastic scintillator plane pairs \\
\# of plate/plane pairs & 282 \\
                               & (121 Calorim. Section + 161 Spectrom. Section) \\
Sampling pitch & 4.54 cm \\
                            & (2.54 cm steel+1.0 cm scint.+1.0 cm air gap) \\  \hline
Active plane segmentation & 96 (or 64) scintillator strips per plane \\
configuration & $\pm 45^o$ w.r.t. vertical, $\perp$ w.r.t. beam axis \\ \hline
Strip width      & 4.1 cm \\
 coating & TiO$_2$ doped polystyrene reflector layer\\
read-out & wavelength-shifter fiber + multi-anode PMT \\  \hline
\hline
\end{tabular}
\caption{Main specifications of the MINOS-ND detector from Reference~\cite{MINOS-ND-tech}.}
\label{tab:MINOS-ND}
\end{table}

The detector is organized in two sections. The upstream fully instrumented section forms the calorimeter, for neutrino interaction vertex definition and neutrino-induced hadronic shower energy measurement. The downstream section is used as a muon spectrometer and only one in every five active planes is instrumented. \\
The MINOS-ND regions, plane configurations, and fiducial volume with respect to the beam can be seen in Figure~\ref{minosregions}.

The MINOS-ND event reconstruction procedure, relevant for the {\sf ArgoNeuT} analysis, uses the topology and timing of the recorded hits from the strips to identify through-going muon tracks originating from CC neutrino interactions in the {\sf ArgoNeuT} liquid argon volume. For typical muon tracks produced by beam $\nu_\mu$ charged-current interactions, the momentum resolution is approximately 5\% using muon range, and 10\% using track curvature.

\section{Commissioning and data taking}
\label{sec:commissioning}
The {\sf ArgoNeuT} detector was commissioned for the physics run in the MINOS-ND hall at Fermilab in spring 2009. After  assembly and cabling, the inner detector was inserted into the cryostat from the front end-cap and then vacuum sealed inside. The argon re-condensation and purification system was also completed, leak tested and connected to the cryostat, as well as the read-out electronics to the TPC wires through connection to the signal feed-through flanges.\\
The commissioning procedure  started with a vacuum pumping phase of the {\sf ArgoNeuT} cryostat vessel, and the connected volumes and transfer lines, down to $\le 9\times 10^{-5}$~mbar. A purging phase with purified argon gas flushing followed by evacuation was then performed in three successive cycles. \\
A liquid argon supply station composed of an array of several parallel 220 L Dewars was connected to the cryostat through a dedicated cryogenic valve system and a filling line that included a set of two filters oriented in series.  
The warm lines and filters were cooled with an initial stream of gas from the liquid argon storage. \\
The filling procedure was then initiated. With a positive pressure inside the vessel and a significant amount of cooling required before liquid began to condense out at the bottom of the cryostat, gas was allowed to vent out of the detector during the fill through the long exhaust pipe to the surface. The entire $\sim$550~lt cryostat was filled in about 12~hours with 6-7 Dewars of liquid argon. The fill was performed intentionally slowly so as to allow the argon sufficient residence time in the inline filters in order to maximally remove the impurities inherent to the commercial product. \\
The cryocooler was enabled during the fill and the liquid argon coming off the recondensation vessel housing the heat exchanger  was directed through the filter manifold of the recirculation system. The fill of the cryostat was considered complete when the capacitive liquid level meter, placed about 30 cm above the top side of the TPC at the neck edging, was covered with liquid. After the fill, the system took a few hours to reach an equilibrium state as all surfaces in contact with argon thermalized at liquid temperature and the cryocooler began re-condensing the boil off gas effectively. Liquid was forced through the filter manifold and the recirculation/purification process was considered stable after this brief post-fill cooling period.\\

HV distribution to the inner detector (cathode, field cage and wire planes) and read-out electronics were activated after steady conditions of the thermal and recirculation parameters were reached. The nominal drift field was smoothly reached by raising the negative voltage on the cathode with a slow HV ramp. 
After signal-to-noise optimization, the final field setting during the physics run was defined as:\\
- Drift field across the TPC volume: $E_d$=481 V/cm (from cathode to innermost Shield plane, $\ell_d=47$ cm).\\
- Field in the first interplane gap: $E_{g1}$=700 V/cm, $r_{T1}=1.45$ (between Shield and Induction plane, $\ell_{g}=0.4$ cm).\\
- Field in the second interplane gap: $E_{g2}$=890 V/cm, $r_{T2}=1.27$ (between Induction and Collection plane, $\ell_{g}=0.4$ cm)\\

The trigger condition for the data acquisition of the physics run is set in coincidence with the NuMI beam spill signal of 0.5 Hz rate. 
At each trigger, the event record  includes a time stamp provided by NuMI accelerator complex. This is used  for 
MINOS-ND and {\sf ArgoNeuT} event matching and tracks association in the event on a spill-by-spill basis.

A first short run period, before the summer beam shutdown, was devoted to detector performance optimization.\\
A measurement of the statistical noise on the wires yielded an RMS of 1.4 ADC counts with negligible coherent noise.
In pure argon, 4 mm of track length for a minimum ionizing particle should have a
mean signal amplitude of $\sim$26 ADC counts in the Collection plane and $\sim$13 ADC counts in the
Induction plane. Thus a signal-to-noise ratio of at least 10 to 1 is expected. 
However, the electron lifetime was not long enough to approach this level.
During the accelerator shutdown in summer 2009, the electron lifetime was improved by continuous GAr purification, reaching $\simeq 750~\mu$s (see Sec.\ref{lifetime}).
  
\begin{figure}[tb]
\begin{centering}
\begin{tabular}{c}
\includegraphics[height=3.in]{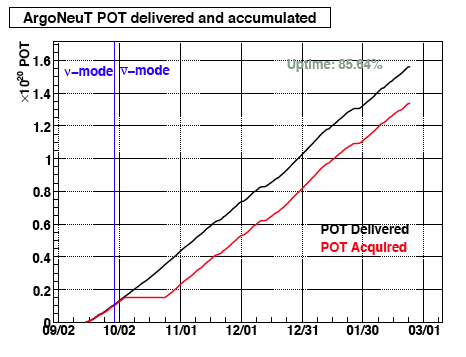} 
\end{tabular}
\caption{The {\sf ArgoNeuT} physics run in terms of delivered/acquired protons on target as a function of date, spanning 2009/2010. The $\sim$2 week downtime in October was due to a failure of a commercial component of the cooling system.}
\label{potplot_uptime}
\end{centering}
\end{figure}

The physics run began in September 2009 and lasted about six months.
From Septmber 2009 to February 2010, the NuMI accelerator delivered $1.335\times10^{20}$ protons on target with the ``low-energy" configuration, of which $8.5\times10^{18}$ in neutrino mode and $1.25\times10^{20}$ in anti-neutrino mode. 
{\sf ArgoNeuT} saw an uptime in terms of ``protons on target (POT) delivered" of about 86\%, including a two-week downtime in October 2009 due to a failure of a commercial component of the cooling system (see figure~\ref{potplot_uptime}). Without including this period of suspended operation, the uptime was about 95\% for the entire physics run. The MINOS-ND was fully operational in coincidence about 90\% of the time.

The POT delivered  to the NuMI beam each spill is provided by the Accelerator Complex, and this information is appended 
to the {\sf ArgoNeuT} event record.\\
The collected events can be classified as :
\begin{enumerate}
\item ``Empty event": the vast majority of NuMI beam spills delivered did not produce an observable interaction within the TPC
as expected due to the very low neutrino cross-section and the limited size of {\sf ArgoNeuT}.
\item ``Through-going track(s) event", where charged particles (mainly muons) produced by neutrino interactions upstream of the {\sf ArgoNeuT} detector 
propagate up to the LArTPC volume. This is the largest sample of events after the "empty event" sample. An analysis of this class of events may be found elsewhere \cite{ArgoNeuT-mu-paper}.  
\item ``Neutrino event" candidate. In this case an interaction vertex with one or two or more tracks has to be present within the LArTPC fiducial volume.
\end{enumerate}
Events not belonging to the above classes include 
 ``Border event" containing a non-certified vertex near the TPC box boundaries, cosmic ray and low energy fully contained tracks or showers induced by neutral punch-through particles from upstream interactions.
 
Events from the  ``through-going tracks" sample were also used for the characterization or monitoring of the main working parameters of the detector during the physics run period, such as the electron drift velocity across the TPC
and the lifetime of the free electron charge in LAr providing daily monitoring of the actual purity of the liquid in the TPC. These topics will be described in  Sec.\ref{vdrift} and Sec.\ref{lifetime} respectively.

 Neutrino events are classified based on a variety of different signatures:
\begin{enumerate} 
\item Charged-current (CC) $\nu_\mu$ candidates with a leading muon track generally escaping the LArTPC volume, 
accompanied by some number of hadrons.
An analysis of this class of events may be found elsewhere \cite{ArgoNeuT-CC-inclusive}.
\item Charged-current $\nu_e$ candidates where an electromagnetic shower develops in the LArTPC volume, starting from the interaction vertex with additional hadron tracks.
\item Neutral-current (NC) candidates  where no muon or electron (electromagnetic shower) are associated with the vertex. This sample includes events with electromagnetic showers from $\gamma$ conversion following the $\pi^o$ decay.
\end{enumerate}

All these provide data samples available for current cross section studies and measurements. 
These consist of approximately 900 $\nu$ CC-interactions in the fiducial volume of the TPC in the neutrino-beam configuration, and 4000 $\nu$ and 3500 $\bar\nu$ CC interactions
in the antineutrino-beam configuration.\\

\subsection{Event Imaging}
\label{EvtImg}

The non-destructive configuration of the wire-planes and the individual wire signal read-out/recording allow for imaging of 
the ionization event in the LArTPC volume.
\begin{figure}[h]
\begin{centering}
\begin{tabular}{c}
\vspace{-0.5cm}
\includegraphics[height=3.8in]{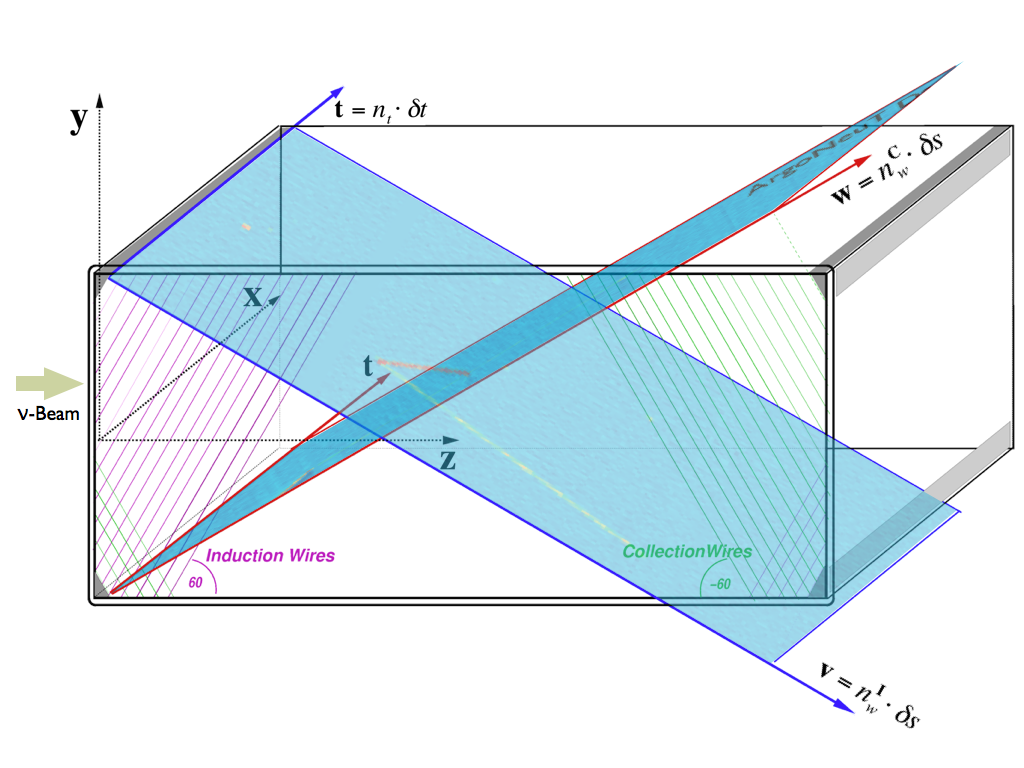} 
\end{tabular}
\caption{Schematic of the {\sf ArgoNeuT} LArTPC and the reference frames adopted for 2D and 3D imaging of the ionization events. 
The coordinates (${\it w,~t}$) for the Collection view and (${\it v,~t}$) for the Induction view are explicitly indicated in terms of 
wire index ($n_W^{I,C}$ and wire pitch ($\delta s$) for the wire coordinate and time tick index ($n_t$) and sampling time ($\delta t$) for the time coordinate.}
\label{wire-chamber}
\end{centering}
\end{figure}

In {\sf ArgoNeuT} each of the two instrumented wire-planes provides a 2D-image corresponding to the event projection on a plane whose axes are identified as ``wire coordinate" and ``time coordinate". 
Both coordinates are discrete, in terms of the wire-number in the plane ($n_w$, from 1 to 240 for both Induction and Collection) and of the time tick of the signal digitization ($n_t$, from 1 to 2048 samples). \\
 A schematic view of the wire plane geometry and of the reference coordinate frames are shown in figure~\ref{wire-chamber}.
 The two projection-planes are indicated as (${\it w,~t}$) for the Collection and (${\it v,~t}$) for the Induction.
The two planes have the time coordinate in common. The wire-coordinates lie along the wire pitch directions.\\
A 2D image of the ionization tracks in the event is obtained for each projection plane. 
Underlying the 2D images are the recorded waveforms $V_{n_t}$ of each wire in the plane, exploited through an accurate signal waveform processing (see Sec.\ref{recon}). The signal pulse
amplitude (i.e. the local ionization charge release) is coded into the color level of the image pixelation (pixel size 
$\delta s\times {\rm v}_d~\delta t~=4\times 0.32~{\rm mm}^2$).\\
Image reconstruction in 3D can be accomplished by combining information from the two 2D views (see Sec.\ref{sec:3Dreco}). The reference system (${\bf x,~y,~z}$) adopted for such spatial reconstruction is also indicated in figure~\ref{wire-chamber}. \\

\begin{figure}[h]
\begin{centering}
\includegraphics[width=5.5in]{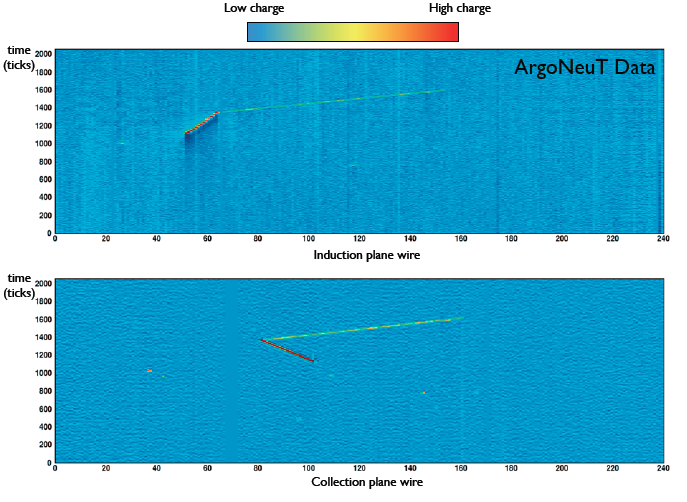}
\caption{A two-prong CC $\nu_\mu$ event candidate (Run \#627, Evt. \#4192):  the yellow trail corresponds to a MIP-like particle escaping {\sf ArgoNeuT} (and reaching MINOS-ND  downstream), the red trail signifies a more densely ionizing particle (presumably a proton, leading to a possible CC-QE signature).}
\label{fig:evt1}
\end{centering}
\end{figure}

\begin{figure}[!h]
\begin{centering}
\includegraphics[width=5.5in]{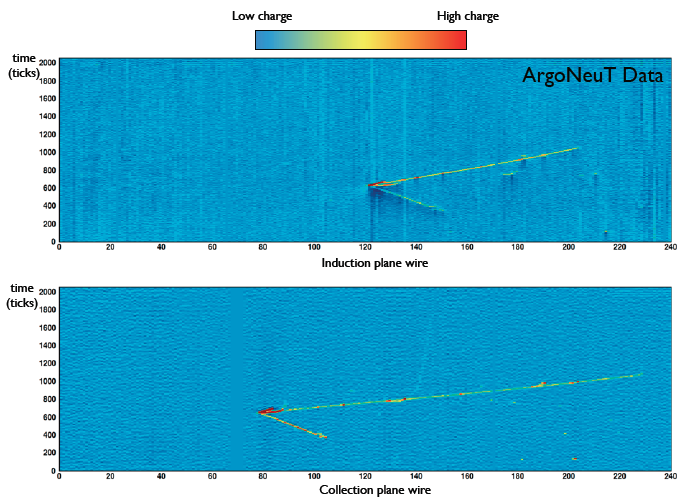}
\caption{A CC $\nu_\mu$-event candidate. Two  short, densely ionizing ($p$-like) tracks are associated with the interaction vertex. The other two $mip$-like tracks escape the detector volume. One is found propagating through MINOS-ND and recognized as a muon, the other is not found in the backing detector and can be associated to a charged pion decaying or interacting in the material between {\sf ArgoNeuT} and MINOS-ND.}
\label{fig:evt2}
\end{centering}
\end{figure}

\begin{figure}[!h]
\begin{centering}
\includegraphics[width=5.5in]{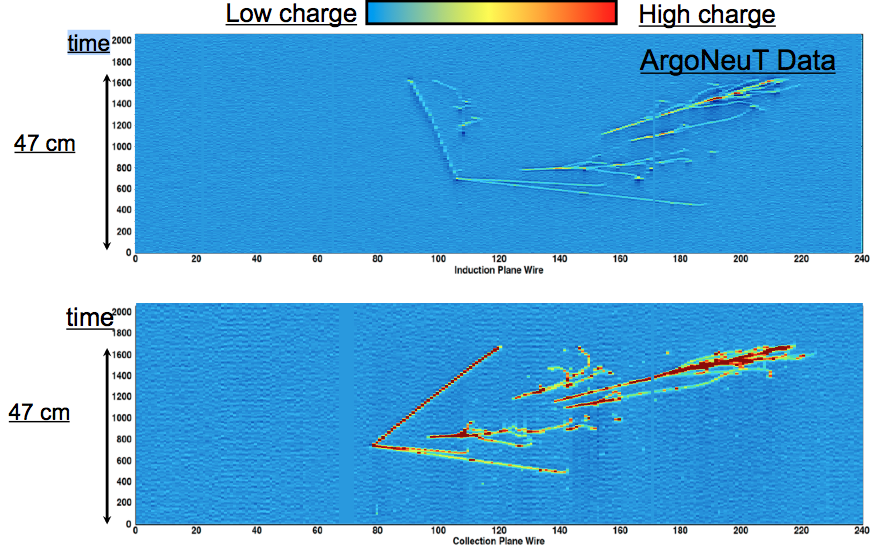}
\caption{Neutrino interaction with 2 prongs at the vertex and four well separated electromagnetic showers propagating through the LAr volume. The showers are visible in both Induction and Collection views.}
\label{fig:evt3}
\end{centering}
\end{figure}

Examples of 2D neutrino event images recorded in the Collection and Induction views are shown
in figure~\ref{fig:evt1}, figure~\ref{fig:evt2} and figure~\ref{fig:evt3}.
In these pictures, the horizontal coordinate corresponds to the wire coordinate in the plane and the vertical coordinate
corresponds to the drift time.  

\section{Event Reconstruction}
\label{recon}
The neutrino event reconstruction procedure combines information separately obtained from {\sf ArgoNeuT} (vertex finding, imaging, energy deposited and particle identification) and MINOS-ND (escaping muon momentum and sign, for the main $\nu_\mu$/$\bar\nu_\mu$ CC sub-sample).

The main steps of the {\sf ArgoNeuT} data processing and matching with MINOS-ND are reported here, while details of the track reconstruction procedure in MINOS-ND can be found in \cite{MINOS-ND-tech}.\\
The off-line software code is organized within the LArSoft package \cite{LArSoft}, a common framework for the simulation and data reconstruction/analysis  of LAr-based experiments at Fermilab.
The general structure of the {\sf ArgoNeuT} off-line event reconstruction chain, presented below, is well established, while the individual steps are subject to continuous improvement. Algorithms and software tools  remain under heavy  development and only a preliminary version of the code has been used up to now.   

As neutrino events in the GeV energy range primarily feature line-like ionization trails from $\mu,~\pi^{\pm}$ or $p$, the event reconstruction development has so far focussed on
straight-line tracks, along with vertex finding algorithms. More complicated pattern recognition software is currently being developed for the reconstruction of  electron tracks and showers, hadron interactions and decays, vertex activity, ... 

At each beam-spill trigger, the {\sf ArgoNeuT} event structure recorded by the DAQ system includes  2$\times$240 digitized signal waveforms from the wires in the Induction and Collection planes.

When a packet of electrons  (e.g. a segment of an ionization track crossing the TPC) is detected by a wire,  
a pulse above baseline is generated within the drift time interval of the recorded waveform from this wire. The shape of the pulse is different for wires in the Induction-plane and in the Collection-plane due to the geometrical and electrical configuration of the TPC planes.
The raw pulse, after signal identification, noise filtering and reshaping, is converted into a "hit" characterized by its peak amplitude 
and coordinates in the wire-time plane. 
 The off-line procedure then uses the hits from both planes to fully reconstruct the ionizing tracks in the event, i.e. the space-coordinates of the hits associated to the track and the energy deposited at those coordinates.

The techniques described here, though developed for the specific task of ionization track reconstruction in LAr,
can be considered common to many other types of applications such as finding signal hits above noise to form a hit-image, clustering proximal hits together, and searching for (while fitting) line-like patterns amongst the clusters.\\

The off-line reconstruction chain consists of the following steps:
\begin{itemize}
\item Raw waveform treatment and noise filtering.
\item Hit construction and identification.
\item Clustering proximal hits.
\item Two dimensional (2D) line reconstruction.
\item Vertex finding.
\item Three dimensional (3D) track reconstruction.
\item Calorimetric reconstruction of deposited energy.
\item Track matching with the MINOS-ND.
\end{itemize}

In the following subsections each of the steps in the reconstruction chain is detailed, and examples of the present algorithms applied to {\sf ArgoNeuT} events are shown.

\subsection{Raw waveform processing: shaping and noise filtering}
\label{raw_wfm}
The first step of the data processing  applies to the digitally recorded raw waveforms ($V_{n_t}$) to obtain the noise filtered, electronics response deconvoluted wire signals ($S_{n_t}$).

\begin{figure}[!h]
\begin{centering}
\includegraphics[width=3.7in]{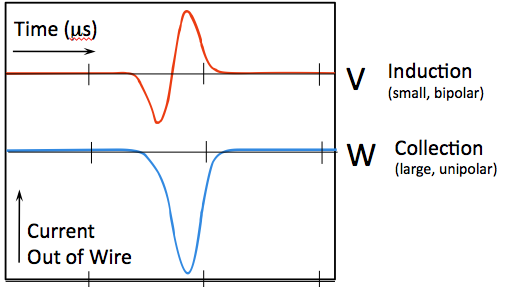}
\includegraphics[width=2.5in]{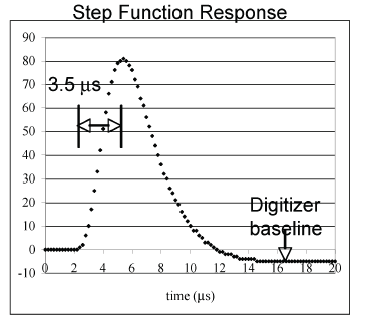}
\caption{[Left]  Current signals for the Induction and Collection wires as induced by an ideal (point-like) drifting charge. 
[Right] The response of the filter and digitizing electronics to a step-function integrator signal, corresponding to a delta function signal on the wire.}
\label{response}
\end{centering}
\end{figure}

In the Induction plane, the current signal from the wires is bipolar in shape, as charges, screened by the Shield plane while drifting in the TPC volume, induce a current only when they cross the Shield plane and move toward, across and away from the Induction plane (figure~\ref{response} [Left-Up]). 
The Collection plane wires yield instead a unipolar current pulse, as after crossing the Induction plane charges move toward up to being collected at the Collection wire (figure~\ref{response} [Left-Down]). \\
At the read-out electronic output the shape of the signal is preserved in both cases due to the features of the narrow Gaussian filter stage (see Sec.\ref{ROelectronics}). In the Collection plane however, output signals are
followed by a negative dip and exponential return to baseline which is caused by the capacitive coupling of the digitizer inputs (see Sec.\ref{ROelectronics}).

\label{hitreco}
 
To decouple and remove the effects of the PFC and ADF stage  electronics, a deconvolution scheme using the Fast Fourier Transform ({\it FFT}) algorithm \cite{sw-tools} is employed. The electronics Response Function of each channel is measured individually from an external test-pulse generator or from the narrowest physical signal detected, as shown with an example in figure~\ref{response} [Right]).
The deconvolution of the electronics Response Function $\mathcal{R}(t)$ from a recorded 
waveform $V(t)$ is performed numerically in the frequency domain to obtain the "true" signal shape $S(t)$. 
The discrete Fourier transforms $\tilde{{\rm v}}_{n_t}$ and $\tilde{r}_{n_t}$ of $V_{n_t}$ and $\mathcal{R}_{n_t}$ respectively are first evaluated 
by the {\it FFT} algorithm. According to the convolution theorem, the true signal $S_{n_t}$ is hence obtained by taking (with the same {\it FFT} algorithm) the inverse Fourier transform of $\tilde{s}_{n_t}$ (=~$\tilde{{\rm v}}_{n_t}/\tilde{r}_{n_t}$) [with ${n_t}$ time tick counter, ${n_t}=1,..,2048$]. 

In the Induction plane, in case of pulses nearly overlapping in time, the bipolar shape of the signal makes this separation difficult. The Induction wire bipolar signals are therefore converted into unipolar shapes. This is also performed by means of a deconvolution in the frequency space, where each Fast Fourier Transformed wire signal is divided by the result of an {\it FFT} transform of the Induction signal shape, the electric field response and a filter that cuts out the low frequency noise.

Filtering of the frequency space is a necessary component of {\it FFT} deconvolution in the presence of noise.  Without low-pass filtering, high frequency noise components are amplified above the signal.  This is handled differently in the two planes due to the differences in the noise and signal shaping.  
In the Collection plane, a technique known as {\it optimal} (or Weiner) filtering \cite{sw-tools} is used, which effectively weights each frequency according to its power in the noise and signal power spectrum.  In the Induction plane a smooth analytic function is used that preserves low frequencies and attenuates high frequencies.

\begin{figure}[h]
\begin{centering}
\includegraphics[width=6.3in]{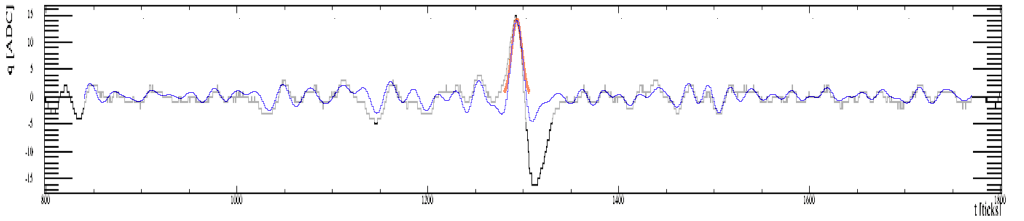}
\includegraphics[width=6.2in]{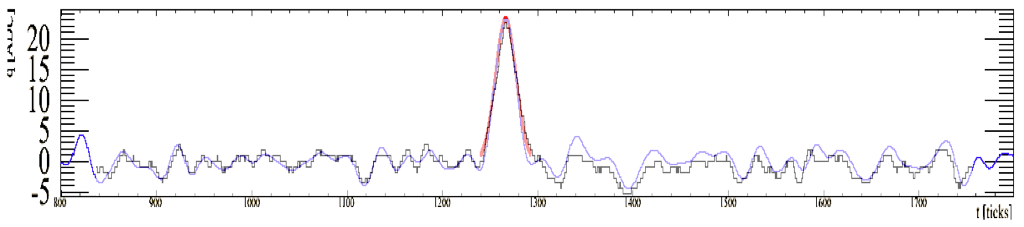}
\caption{Real data (crossing muon track parallel to the wire planes):
wire raw waveform (black) and deconvoluted waveform (blue) with  gaussian fit superimposed (red). [Top] Induction wire hit parameters: 
 wire number $n_w^I=137$, time position $n_t=1290$, amplitude $A \simeq 14$ ADC, width $\sigma \simeq 6$ time ticks. 
[Bottom]  Collection hit parameters:  wire number $n_w^C=127$,  time position $n_t=1264$, amplitude $A \simeq 25$ ADC, width $\sigma \simeq 9$ time ticks.}
\label{fig:Raw-Filt-Sign}
\end{centering}
\end{figure} 
The {\it FFT} output waveforms $S_{n_t}$ feature smooth and unipolar pulses from the wires in both planes. The shape of the pulses turns out almost 
symmetric, with similar rise and fall times. A gaussian-shape approximation for the pulse allows a simplified approach to identifying the interesting regions of the waveform which are referred to as ``hits".  \\
An example is given in figure~\ref{fig:Raw-Filt-Sign}.  A track segment of a muon crossing the TPC parallel to the wire planes induces the bipolar pulse (black histogram) in the raw waveform $V_{n_t}$ from an Induction wire [Top] and a unipolar pulse in the Collection wire [Bottom].
The deconvoluted waveform $S_{n_t}$ (blue line) is superimposed, with the hit (red gaussian fit) identifying the muon signal.

\subsection{Hit Identification}
\label{sec:HitID}
The hit-finding algorithm scans the  processed wire waveform looking for local minima.  If a minimum is found, the algorithm follows the waveform after this point until it finds a local maximum.  If the maximum is above a specified threshold, the program scans to the next local minimum and identifies this region as a hit.  If the local maximum is below threshold, it rejects that region and scans to the next local minimum to begin again.\\
Once one or more hit regions of the waveform are identified, they are each fit with a Gaussian function whose features identify the correct position (time coordinate), width and height of the hit.  \\
If multiple hit regions are consecutive, and the region between them is above a threshold, then multiple hit fitting is used to extract the parameters of the individual hits.  The built-in histogram fitting algorithm embedded in the {\it ROOT} software package ~\cite{sw-tools} is used. 
This algorithm requires seeding initial values to the parameters reasonably close to the correct ones in order for the fit to converge.  
\begin{figure}[h]
\begin{centering}
\includegraphics[width=6.3in]{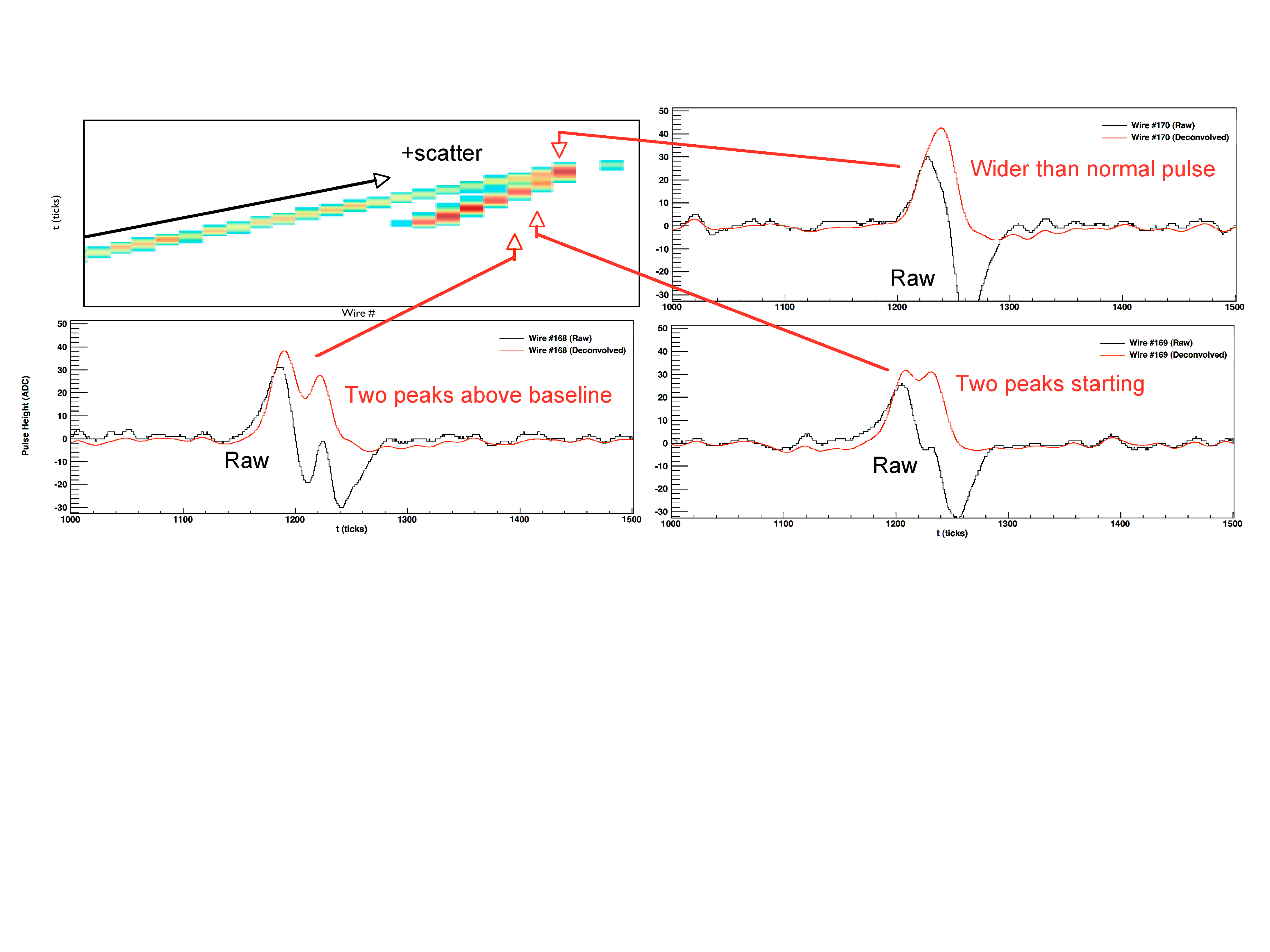}
\vspace{-5.1cm}
\caption{(Upper left) A set of tracks as seen on the (deconvoluted) induction plane. The wire views on three adjacent wires are also shown in order to demonstrate the effects of deconvolution on the raw wire pulses. The raw data can be seen in black and the deconvoluted data can be seen in red.
As effects of the deconvolution algorithm, signal-to-noise increases, effects of adjacent signals on each other are removed, and the intrinsic bipolar shape on the Induction plane signal is converted into a unipolar one.}
\label{decon_ind}
\end{centering}
\end{figure} 
The initial half-width value is given as a parameter for the Induction and Collection planes separately, the values of which are determined from typical single hit widths in the data, these are $6.0$ and $7.8$ time ticks, respectively.  
The initial positions are given as the local maximum positions. The initial signal amplitude for single hits is simply the height of the maximum.

 For the peak height of multiple hits a more complex procedure is required due to added signal from nearby hits. In these cases,
to get the initial values for the individual hits a simple linear approximation is made.  The positions and widths are assumed to be correct.  Then, for a chain of multiple hits, the measured height at one peak is given by its height plus the contributions of its neighbors.  Then the measured height of peak $i$, $A^*_{i}$ is given by $ A^*_i= \sum A_j~f(t_j-t_i;w) $,
  where $f(t;w)$  is the normalized model signal shape,  $w$ its width, $t_{i}$ the center position of peak $i$, and $A_{i}$ is the approximate signal height of hit $i$ alone.  It is easily seen that this forms a linear equation $\vec{A^*} = M \vec{A}$, where $\vec{A^*}$ and $\vec{A}$ are the observed and approximate individual hit amplitudes, and {\it M} is formed by the values of the normalized model function for each hit evaluated at the point of the other hits.  This matrix is symmetric with unity on the diagonal leaving less than half of the matrix elements to compute.  Solving this system then gives a good approximation of the initial signal amplitudes.  This is done using the {\it DecompSVD} method within {\it ROOT} \cite{sw-tools}.
  
The hit finding procedure scans all deconvoluted waveforms from both planes. 
 Hit parameters, wire number and position (time tick), amplitude and width determined from the fit, are recorded.  The hit start and end times are defined as the center position minus and plus the width, respectively.  \\
Detected hits, with amplitude above selectable threshold, can be displayed in the corresponding wire-time plane (${\bf w,~t}$) for the Collection and (${\bf v,~t}$) for the Induction (Sec.\ref{EvtImg}). The ADC-scale of the hit amplitude is converted into a color-level scale to visually indicate
low (yellow) or high (red) local ionization density.
This is shown in figure~\ref{decon_ind}
where hits of a two-pronged track are seen in the Induction view [Top-Left], along with the raw and deconvoluted signals from three adjacent wires.

\subsection{Hit clustering}
\label{sec:hitClust}
Hit variables from the hit identification process are passed from each view along to subsequent reconstruction methods that cluster the hits together in the same wire-time plane. Line-like clusters will be then recognized as 2D-projections of ionization trails in the LArTPC volume.\\

{\it Density-Based Spatial Clustering}\\
The hits are first clustered together through an algorithm based on {\it DBSCAN} (``Density-Based Spatial Clustering of Applications with Noise") ~\cite{sw-tools}, which uses the density of hits as a distinguishing characteristic.\\
The {\it DBSCAN} implementation in the {\sf ArgoNeuT} hit-clustering code features an elliptically-shaped neighborhood in the wire-time space, with two parameters (the semi-major/minor axes of the ellipse, $\epsilon_{1}$ and $\epsilon_{2}$) to take into account different resolutions, along the wire and time axes. \\
In order to form a cluster, the algorithm starts with an arbitrary $h$-th hit-point  among those found in the current view and retrieves all points that are ``density-reachable" in consideration of the specified parameters. 
A point $q$, which represents the central coordinates of a hit in the wire-time space,  is ``directly density-reachable" from a point $p$ if it is not farther away than a given distance $\epsilon$  (i.e. it is part of its predefined $\epsilon$-neighborhood), and if $p$ is surrounded by sufficiently many points. 
The point $q$ is ``density-reachable" from $p$ if there is a sequence $p_1,\ldots, p_n$  of points, with $p_1 = p$  and $p_n = q$,  where each $p_{i + 1}$  is directly density-reachable from $p_i$. The definition of density-reachable is not symmetric since $q$ might lie on the edge of a cluster, having too few neighbors to count as a genuine cluster element.  \\
A cluster, which is a subset of points within the data set, satisfies two properties: (1) all points within the cluster are ``density connected" (i.e. two points $p$ and $q$  are density-connected if there is a point $s$ such that $s$ and $p$ as well as $s$ and $q$ are density-reachable) and (2) if a point is density-reachable from any point of the cluster, it is part of the cluster as well. 
Core points are those points that are inside of a cluster and border points are those which are on the border of a cluster. \\
A cluster is started if $h$ is a core point. If $h$ is a border point, no points are density-reachable from $h$ and {\it DBSCAN} continues on to the next point.
\begin{figure}[h]
\begin{centering}
\includegraphics[width=4.0in]{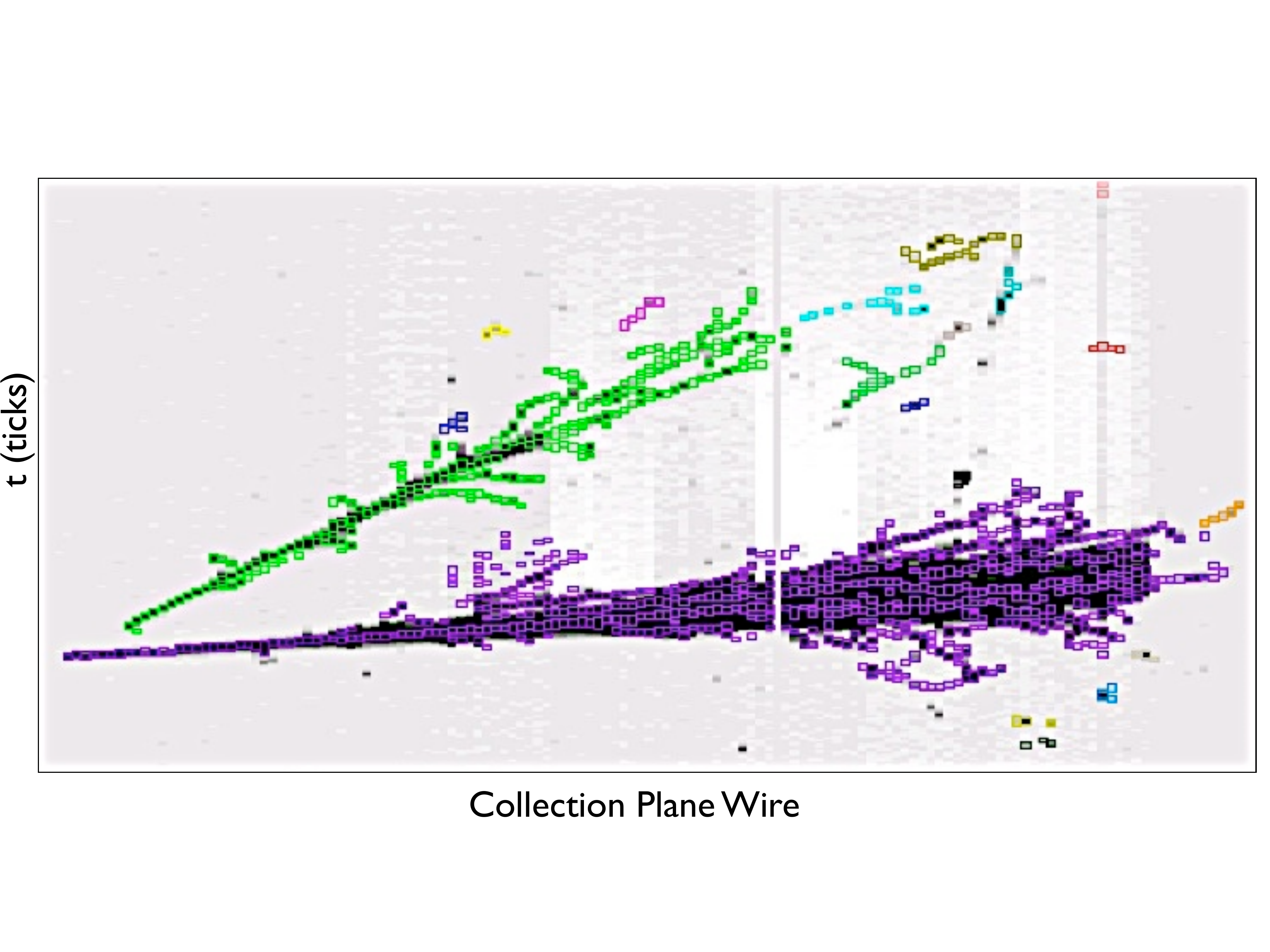}
\vspace{-1.0cm}
\caption{An example of 2D hit-clustering from {\it DBSCAN} applied on real events. Two main clusters corresponding to two electromagnetic showers are identified by different colors.}
\label{fig:dbscan-shwr}
\end{centering}
\end{figure}

An examples of 2D hit clustering using the {\it DBSCAN} algorithm on {\sf ArgoNeuT} data can be seen in figure~\ref{fig:dbscan-shwr}.
The algorithm is seen to nicely define and separate two main gamma-induced electromagnetic showers into separate clusters found in a neutrino candidate event. When tracks originate from a common vertex and are thus connected there
a further reconstruction algorithm (Hough transform) is applied to fit multiple lines to these clusters since they may contain multiple tracks. \\

{\it Hough Transform line reconstruction}\\
A Hough transform \cite{sw-tools} is used in the reconstruction software to identify all line-like objects contained in a given {\it DBSCAN} cluster.  
The Hough transform creates a parameter space, or {\it accumulator}, filled according to the hit locations within a {\it DBSCAN} cluster.  A straight line can be parameterized by $r=x\mathrm{cos}(\theta)+y\mathrm{sin}(\theta)$, which represents a curve in the $(r,\theta)$ plane that is unique to each $(x,y)$ point. If the curves corresponding to two points are overlaid, the $(r,\theta)$ crossing point represents the parameters of a line that traverses both points. The transform employed in the reconstruction takes advantage of this fact, parameterizing each of the hits (wire,time) in a {\it DBSCAN} cluster and placing them in a Hough accumulator $(r,\theta)$ space.\\
The Hough accumulator's $r$ and $\theta$ coordinates are discretized in the form of cells of finite height and width \cite{sw-tools}, and the number of these accumulator cells determine the algorithm's resolution.   The dimensions of the accumulator are chosen in consideration of resolution, processing time, and accumulator signal-to-noise ratio.  After all of the image points have been positioned in the discretized Hough accumulator, the cells with weight above some threshold are selected as line candidates. The $n$ mutually exclusive cells with the largest weight can be considered lines in the case that the algorithm is searching for $n$ lines.

After all of the hits have been added to the Hough accumulator, the cell with the largest weight (i.e. curve-crossings) is found. The center-of-mass of the 3$\times$3 cell window with the weightiest cell at center corresponds to a candidate line. The center-of-mass of the 9-cell system is used to form the candidate line, instead of the actual central cell coordinate, so that the extracted line parameters are not limited by the cell coordinate's discreteness.
The process is iterated, considering only hits that have not yet been associated with a line (via a non-maximal suppression technique), until the ``minimum weight to be considered a line" (threshold) and/or ``maximum number of lines to be found in a cluster" has been reached.

\begin{figure}[h]
\begin{centering}
\includegraphics[width=4.in]{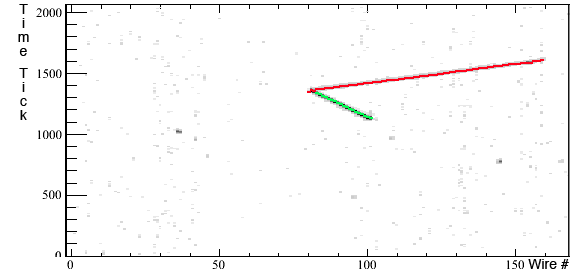}
\vspace{-0.5cm}
\caption{2D hit-clustering with Hough Transform line reconstruction 
(Run\#627, Evt.\#4192, shown in figure~\ref{fig:evt1}). The gray scale displays reconstructed hits while colors are used to indicate the two lines found with the Hough line-finding algorithm. The two lines correspond to the two prongs in this event.}
\label{fig:huogh} 
\end{centering}
\end{figure}
Multiple accumulator convergences are seen in the case of multiple line candidates.
The entire process is repeated for all clusters in the event. After all of the image points have been positioned in the discretized Hough accumulator, the cells with weight above some threshold are selected as line candidates. 
Before finding the largest cells in the accumulator, the algorithm can optionally smooth the accumulator with a 2D Gaussian convolution in order to reduce the effect of hits which do not fit well. 
Figure~\ref{fig:huogh} shows a two prongs neutrino event and the results of the Hough line algorithm applied to it. 
The Hough line-finding algorithm properly separates them into two lines corresponding to the two tracks from the interaction vertex.

The Hough line-finding algorithm can sometimes break a single line into multiple line components of slightly varying slope and intercept if there are more $(r,\theta)$ cells to choose from the Hough accumulator.  Lines with similar slope and connected endpoints are merged together on a plane-by-plane basis after performing the line-finding procedure in preparation for 3D track matching.

\subsection{Interaction Vertex finding}
\label{sec:VtxFind}
The two dimensional neutrino interaction vertex is found using the 2D line-like clusters in each plane view. 

Identification of the vertex proceeds as follows. The line-like {\sf ArgoNeuT} cluster which is matched to a MINOS track or otherwise (in case of no match with MINOS) the longest line-like cluster is identified. The start point of this cluster is a guessed vertex in 2D. All other line-like clusters in the plane are examined. All the line-like cluster in the plane whose point of closest approach to the guessed vertex has a chi-square less than some minimum tolerance are associated to the vertex, where the errors in the denominator of each contribution to the $\chi^2$ come from the errors on the calculated 2D hit widths.
Delta rays can interfere with this simple vertex finding algorithm and there is a protection to prevent the point where the delta ray diverges from its ÒparentÓ muon from being identified as the interaction vertex. Basically, clusters are identified as delta rays if they have few hits, a similar slope as the longest cluster in the track, and are far away from the longest clusterÕs start point. After the vertices are identified in each plane, they are matched in terms of a common drift time and the three dimensional vertex is determined.

\subsection{3D track reconstruction}
\label{sec:3Dreco}
Tracks are reconstructed in 3D (${\it x,y,z}$ space coordinates) by combining associated 2D line-like 
clusters that are identified in both views of the TPC (${\it v,t}$ coordinates for the Induction plane and ${\it w,t}$ for the Collection plane). Among the adopted association criteria, the main one requires the line-clusters to have endpoints with same
drift coordinates in the two views.
The common time is defined within a time window tolerance, 
after accounting for the drift distance between the Induction and Collection planes.
\begin{figure}[!h]
\begin{centering}
\begin{tabular}{c}
\includegraphics[height=4.0in]{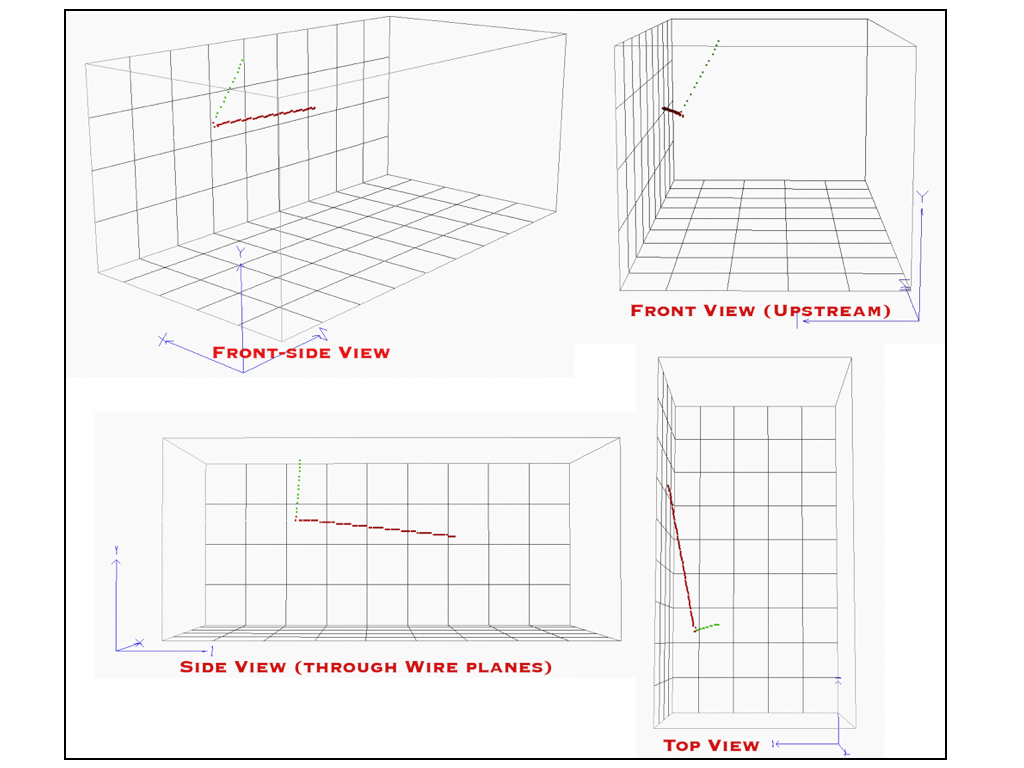}
\end{tabular}
\caption{Neutrino event reconstructed in 3D space (Run\#627, Evt.\#4192, 2D views shown in figure~\ref{fig:evt1}). One track exits the TPC volume through the cathode plane and propagate towards MINOS-ND.}
\label{event3D}
\end{centering}
\end{figure}

Geometrical parameters of the track in space such as the directional cosines and the ``track
pitch length", i.e. the effective length of the portion of track exposed to a single wire (depending on
the orientation of the track with respect to the direction of the wires in the plane), are reconstructed.
The track pitch length from the 3D reconstruction is necessary information for the calorimetric measurement (as shown in the next section). 
Once a 3D track is identified, a hit-by-hit association procedure is applied to match hits from the
two wire planes to obtain a fine-grained 3D image of the event (see figure~\ref{event3D}). This last step is based on a dedicated matching algorithm to form 3D points. This algorithm can also be applied to the 3D reconstruction of arbitrary trajectory tracks that are not necessarily straight lines.

Space points associated with track hits  refer to a cartesian reference system with ${\it x,~y,~z}$ coordinates related to the wire (${\it v}$ for Induction, ${\it w}$ for Collection) and drift-time (${\it t}$) coordinates through the following relations (see figure~\ref{wire-chamber} for indication of the reference systems in use):
\begin{eqnarray}
{\it x}~=~{\it t}~{\rm v}_d\\
{\it y}~=~\frac{{\it w-v}}{2~\cos\alpha}\\
{\it z}~=~\frac{{\it w+v}}{2~\sin\alpha}~-~\frac{{\rm h}}{2~\tan\alpha}
\end{eqnarray}
where v$_d$ is the drift velocity in the TPC volume (see Section~\ref{vdrift}), $\alpha$ is the absolute value of the orientation angle of the wires in the 
wire-planes ($|\pm60^o|$ with respect to the horizontal, neutrino beam direction) and $h$ is the TPC box's height.\\
The drift time is corrected for pre-sampling as well as offsets due to the longer drift distance to the Collection plane.

\subsection{Calorimetric energy reconstruction and particle identification}
\label{sec:calorimetry}
The energy loss along the path ($dE/dx$) of charged particles traversing a volume of liquid argon depends on the particle energy and type.\\
Energy is deposited by ionization at a rate of 23.6 eV per ($e^-$-Ion$^+$) pair (the $W_e$ value in LAr) \cite{miyajima}. 
 However,  only a fraction of the released charge survives fast recombination occurring before the opposite charges are spatially separated under the electric field action \cite{jaffe}. This effect (often indicated as ``charge quenching") and correspondingly the surviving ``free charge" fraction depend on the local ionization density (the higher $dE/dx$, the lower is the free charge fraction). 
Free electrons forming the track of the particle trail are then uniformly transported along the field lines onto the anode plane.
A further loss occurs during the drift due to electron attachment to residual electro-negative impurities diluted in LAr. 
At the wire planes, each segment of the ionization track is detected by a pair of wires (in the Induction and the Collection plane) 
and the corresponding hits are recorded.\\
The algorithm for the deposited energy reconstruction in the present off-line code was optimized for line-like tracks 
and is being extended to more complicated track patterns (e.g. electron tracks or showers).\\
The relevant quantities for calorimetry are obtained from the 2D- and 3D-track reconstruction.  
These are the hit amplitude, the hit time coordinate and the track pitch length $\delta x$ associated to the hit wire, for all the hits belonging to the track. \\
The hit amplitude (in ADC counts), measured in the Collection plane because it provides higher gain, corresponds to the charge $Q_{det}$ (in fC units) in the track pitch detected on the wire (for the ADC to fC conversion factor, see Sec.\ref{ROelectronics}). \\
To account for the charge loss along the drift due to impurities, a first correction is applied to obtain the free charge after recombination 
$Q_{free}=Q_{det}/e^{-t/\tau_e}$, where $t$ is the hit time (presampling subtracted) and $\tau_e$ is the current electron lifetime, routinely measured in {\sf ArgoNeuT} during the physics run (see Sec.\ref{lifetime}).\\
Finally, to account for the charge loss due to recombination, a second correction is applied to obtain the total charge released $Q_{0}=Q_{free}/R$.
The recombination factor $R$ is derived from a parameterization of the quenching effect in LAr reported in \cite{sala} and based on the semi-empirical Birks's model developed for the description of quenching effects in scintillators  \cite{birks}. 
The $R$ factor is a non-linear function of the ionization density $(dQ_{free}/dx)$ freed at the actual electric field strength.
The free ionization density along the track can be sampled by the hit amplitude to the track pitch length ratio $(Q_{free}/\delta x)$, which allows calculating  the value of the $R$ factor.\\
The charge $Q_0$ released in the track pitch is directly related to the energy deposited ($E=W_e Q_0$).  
The energy loss along the track ($dE/dx$) can thus be estimated in steps of length $\delta x$, and the total 
energy deposited along the track is obtained by summing over the steps. \\

\begin{figure}[!h]
\begin{centering}
\vspace{-0.3cm}
\begin{tabular}{c}
\includegraphics[height=2.3in]{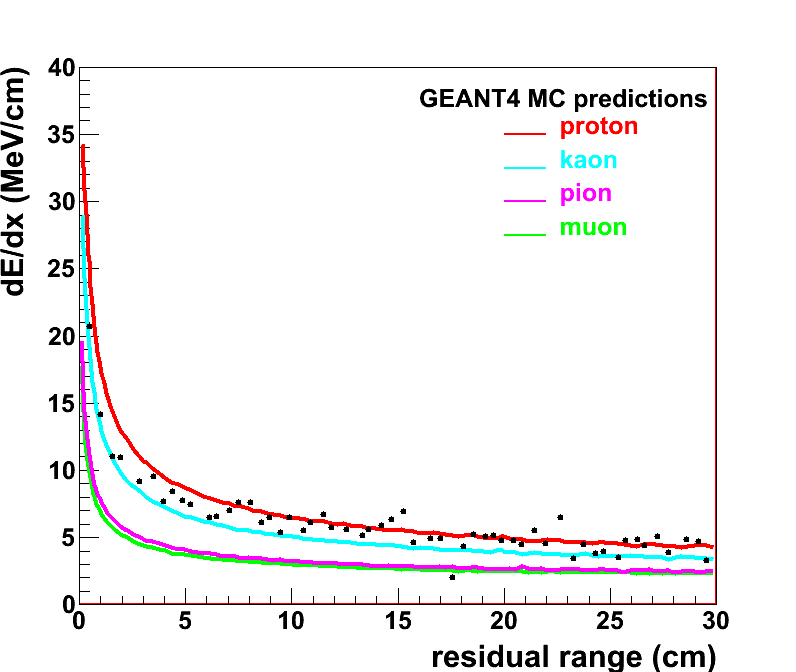}
\includegraphics[height=2.3in]{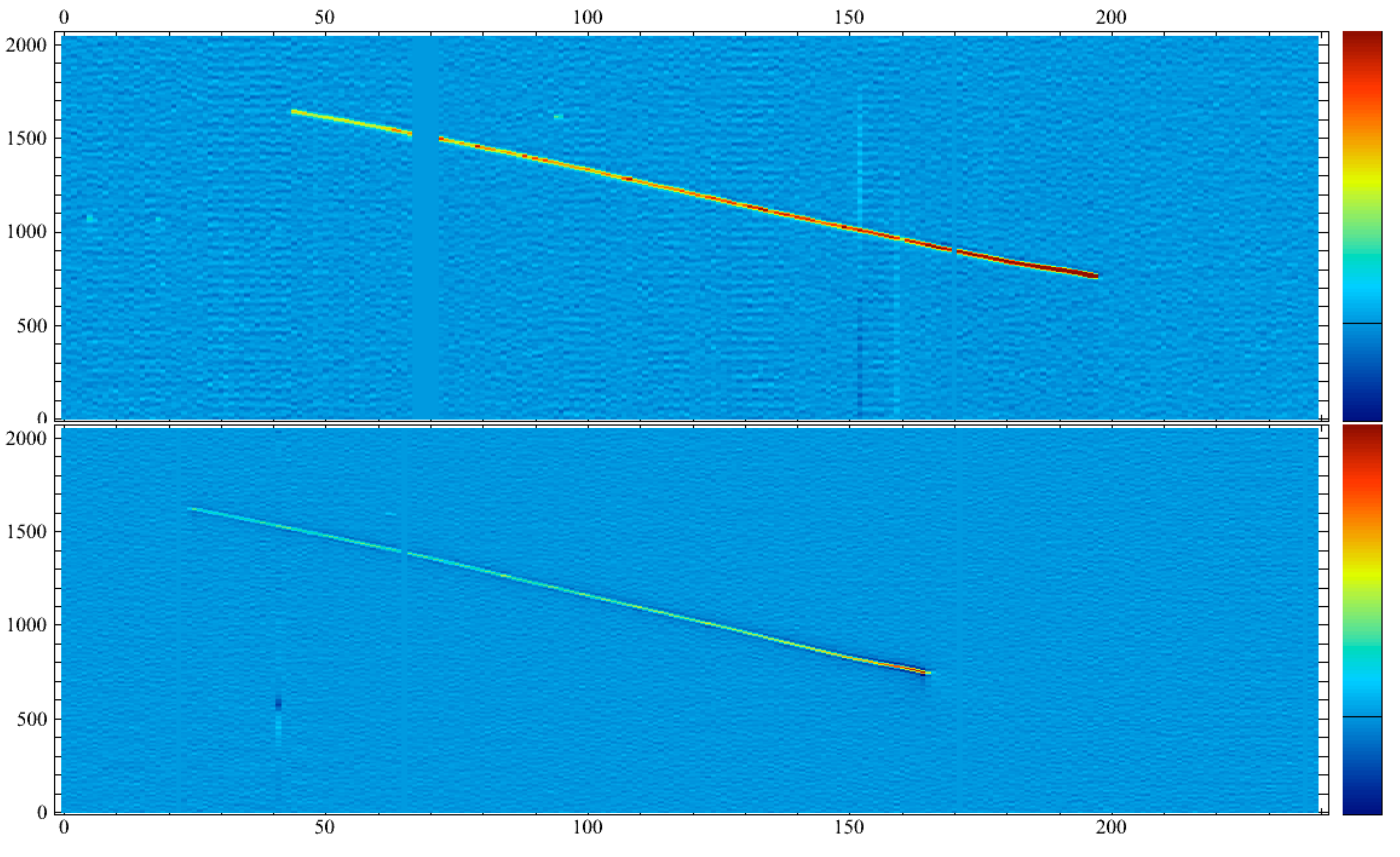}
\end{tabular}
\caption{[Left] Energy loss per unit track length (average value) as a function of residual range (distance to the track end) for different charged particles.
Experimental data from calorimetric reconstruction of the energy loss along the stopping track shown on the right are superimposed. [Right] Evt.\#6474, Run\#650: 2D views from Collection Plane (top) and Induction Plane (bottom).}
\label{fig:PId}
\end{centering}
\end{figure}
If the incident particle slows down and stops in the LArTPC active volume,
the energy loss displayed as a function of the residual range (the path length to the end point of the track)
is used as a powerful method for particle identification (PId).
Charged particles of different mass (or charge) have in fact different increasing stopping power at decreasing distance 
from the track end, as shown in figure~\ref{fig:PId} [Left], where the different curves come from simulation of stopping muons, pions, 
kaons and protons in LAr. \\
As an example of application of the current calorimetric reconstruction and particle identification algorithms, 
the energy loss along a recorded stopping track, shown in figure~\ref{fig:PId} [Right], as sampled by 147 wires in the Collection plane, 
is reported (black dots) on the $dE/dx$ vs. residual range plane [Left]. 
The distribution of the experimental points agrees with the proton hypothesis. 
For a measured track length of 74.1 cm in LAr, 
the incident kinetic energy of a stopping proton also agrees with the total deposited energy of 352.3 MeV 
from the calorimetric reconstruction.

\subsection{MINOS-ND Track reconstruction and association}
\label{sec:minos-assoc}
\begin{figure}[!h]
\begin{centering}
\begin{tabular}{c}
\includegraphics[height=2.7in]{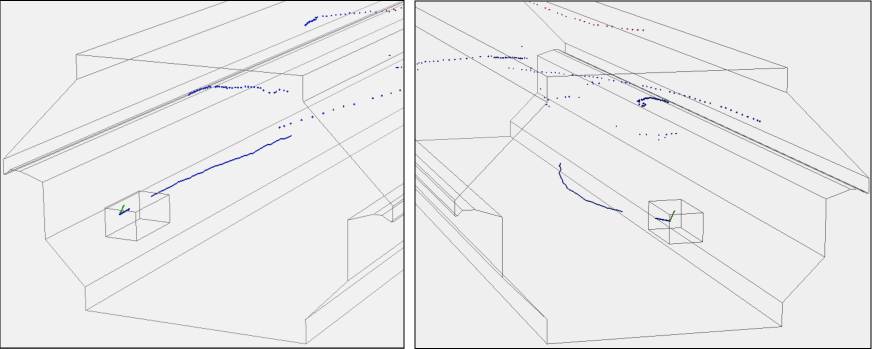}
\end{tabular}
\caption{Full neutrino event reconstruction with 3D {\sf ArgoNeuT}-MINOS ND track matching (Run\#627, Evt.\#4192). This event was already shown in previous figures
at different stages of the reconstruction procedure, from 2D imaging to hit clustering up to 3D display).}
\label{argnt-mns}
\end{centering}
\end{figure}

Particles from GeV-neutrino interactions in {\sf ArgoNeuT} can easily propagate outside the TPC boundaries, and are thus
identified as exiting tracks. In particular, energetic muons can easily reach and be detected as entering tracks 
in the downstream MINOS-ND, as shown in figure~\ref{argnt-mns}.   \\

Track reconstruction in MINOS-ND is performed by MINOS off-line analysis code \cite{MINOS-ND-tech} and the results are provided directly by the MINOS experiment.  A track-finding algorithm is applied. It uses a Hough-transform, embedded in a Kalman filter algorithm, to identify the initial track seed. 
Track segments are then chained together to form longer tracks taking into account timing and spatial correlations. The track momentum is estimated from range if the track stops within the detector, or from a measurement of its curvature in the MINOS-ND toroidal magnetic field if it exits. The curvature measurement is obtained from fitting the trajectory of the track using Kalman filter techniques that take into account bending of the track from both multiple Coulomb scattering and the magnetic field. This procedure also determines the charge of the reconstructed track.\\
All MINOS-ND tracks with first hit coordinate along the beam axis $Z\le$ 20 cm from
the first MINOS-ND plane ($Z$=0) are considered as entering tracks and candidate for {\sf ArgoNeuT}-MINOS
matching. 

Tracks exiting {\sf ArgoNeuT} and tracks enteringMINOS-ND are pre-selected for matching on a spill-by-spill basis based on a common timestamp from the accelerator complex.\\
The actual matching is based on the orientation and position of the {\sf ArgoNeuT} and MINOS-ND tracks relative to one another. 
Two alignment requirements are imposed: (1) on the radial difference between the
exiting {\sf ArgoNeuT} track projected onto MINOS-ND first plane and the MINOS-ND entering track vertex, $\Delta r\le 25$ cm and
(2) on the angle between the {\sf ArgoNeuT} track direction at TPC exit and the MINOS-ND track at vertex, cos$\theta\le$ 0.4 rad.\\
A best alignment criterium is adopted when multiple MINOS-ND tracks and/or multiple {\sf ArgoNeuT} tracks occur:
the {\sf ArgoNeuT}-MINOS track pair combination that minimizes the quantity $\Delta r/$cos$\theta$ is chosen as the correct
 matched track combination.

\section{Detector performance}
\label{sec:det-perf}
A large sample of events from the through-going muon sample, likely generated by $\nu$-interactions in the upstream material and collected all throughout the {\sf ArgoNeuT} physics run period, have been used for both characterization and continuous monitoring of the main working parameters of the detector.  \\
{\sf ArgoNeuT}'s automated reconstruction software is employed in order to find and characterize through-going tracks and the hits associated with them.

\subsection{Electron drift velocity}
\label{vdrift}
The  drift velocity of free electrons in liquid argon (v$_d$) is an 
important parameter. It is used to calculate the drift 
coordinate necessary for a complete three dimensional 
reconstruction of the tracks in the event images.\\
The drift velocity depends on the value of electric field applied in the LAr active volume and on the temperature of the liquid.\\
The electric field throughout the TPC drift region is set at $E_d$=481~V/cm.
The field dependence of the drift velocity is usually expressed as v$_d=\mu~E_d$, i.e. it is proportional to the field 
through the electron mobility ($\mu$, to first approximation a constant in LAr in the $\sim$~0.5 to 1 kV/cm EF range).\\
The temperature of the liquid is 88.4$\pm$0.1 K. This is determined by the gas pressure above the liquid controlled and actively adjusted around the set point value of 2.0$\pm$0.2~psig (+138 mbar relative to atmosphere at $\sim$1003 mbar absolute pressure at the detector location, about 80 m above sea level).
The temperature probes located in different positions inside the cryostat confirm this value within 0.3 K of systematic precision. \\

Direct v$_d$ measurement and comparison with expectation provide important information on the correct 
operation of the LArTPC. \\
Drift velocity measurements can be performed on a single event basis by exploiting ionization tracks crossing the entire drift distance of the detector.
The time difference between the hits, in the waveforms from the two wires detecting the entry point and the exit point of the track, corresponds to the drift time from the cathode to the anode wire plane. The drift velocity is thus given by the drift distance-to-time ratio. \\
Tracks at large incident angle with respect to the beam direction to cross the entire drift distance are however very unlikely to be found in the recorded through-going muon sample (mainly from neutrino interactions upstream of {\sf ArgoNeuT}).

The drift time from cathode to wire plane can alternatively be determined using through-going tracks that cross just one of the two planes, either the cathode or the anode wire plane, and measuring their entry time or exit time. \\
The shield plane, delimiting the TPC drift volume opposite to the cathode, is not instrumented for signal read-out. Therefore, the successive Induction plane has to be used for drift time determination. The corresponding drift distance includes the nominal drift length of the detector ($\ell_d$=470 mm from cathode  to Shield plane at $E_d$) together with the gap between Shield and Induction plane at a higher field value ($\ell_g$=4 mm at  $E_{g1}= r_{T1} ~E_d$, with $r_{T1} $=1.45). A correction for the field change in the gap will be taken into account in the drift velocity determination, as well as a correction due to thermal contraction of the TPC frame from room to LAr temperature in the drift direction.
 \begin{figure}[!h]
\begin{centering}
\includegraphics[height=2.3in]{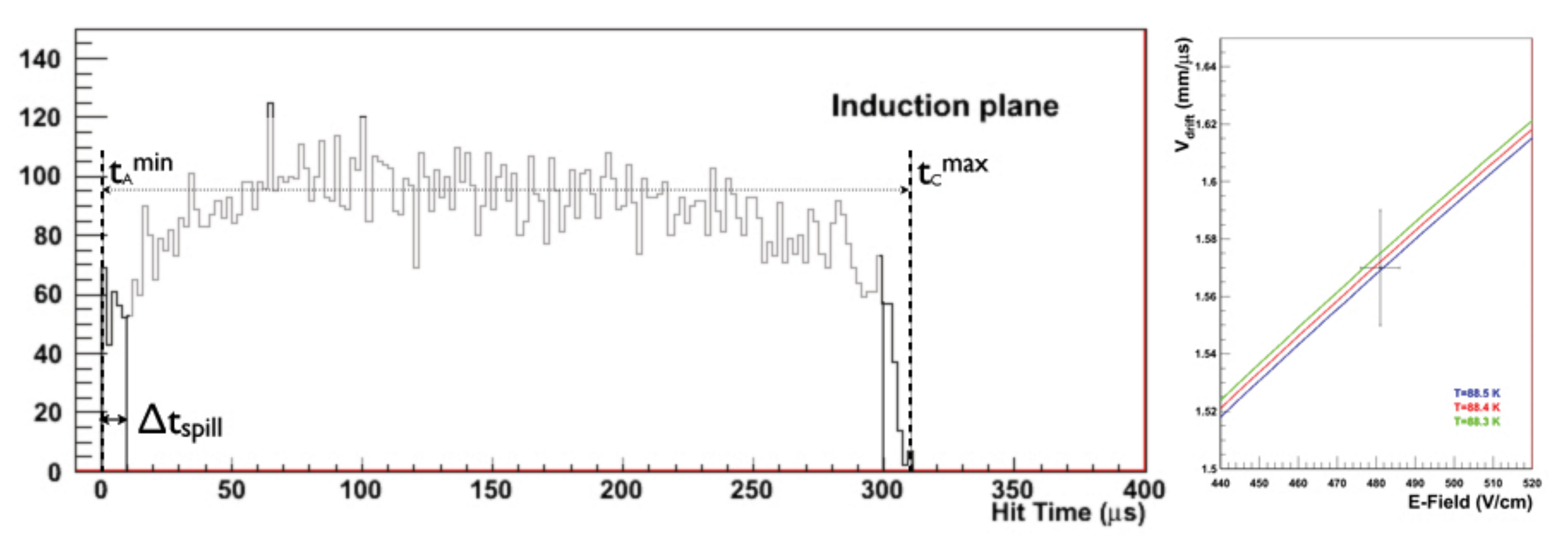}
\caption{[Left]  Hit time distribution from the Induction plane for neutrino induced through-going muon tracks. The sample includes tracks crossing either the cathode or the Induction plane of the TPC. These tracks provide the hits 
at the edges of the distribution and thus determine the drift time measurement through Eq.\ref{eq:t_d}. The precision on the determination of the r.h.s. edge of the distribution (cathode maximum crossing time detected, $t_{C}^{max}=310\pm 2 \mu$s) is the main contribution to the systematic error of the drift velocity measurement (Eq.\ref{eq:v_d}).
[Right] The drift velocity measurement in {\sf ArgoNeuT}, and associated systematic uncertainties, is compared with expectation values from ~\cite{walkowiak} as a function of the electric field and for different LAr temperatures corresponding to the range of variation of the current LAr temperature during operations. }
\label{drift-time}
\end{centering}
\end{figure}

The distribution of the reconstructed hit times in the Induction wire signals from a sample of neutrino induced through-going muon tracks is shown in figure~\ref{drift-time} [Left]. A fraction of these tracks cross either the cathode or the anode plane of the TPC. \\
As neutrino interactions are randomly distributed in time within the NuMI spill duration ($\Delta t_{spill} = 9.7~\mu$s),
 the absolute hit time corresponding to a single plane crossing is also distributed over this time interval.
The earliest neutrino interactions, at the beginning of the spill, will produce tracks with the minimum value of the absolute hit time ($t_{I}^{min}$) when crossing the Induction plane, while the latest neutrinos in the spill 
produce tracks with maximum hit time ($t_{C}^{max}$) when crossing the cathode.\\
 The drift time $t_d$ from cathode to Induction plane can thus be determined from the hit time distribution 
 of through-going tracks as
 \begin{equation}
t_d~=~t_{C}^{max}-t_{I}^{min}-\Delta t_{spill}~=~300.5~\mu {\rm s}
\label{eq:t_d}
\end{equation}
where the values are obtained from figure~\ref{drift-time} [Left]. 

The electron drift velocity can finally be calculated as:
\begin{equation}
{{\rm v}_{d}}~=~\frac{\ell_d~+\ell_g/r_{T1}~-\Delta\ell}{t_d}~=1.57 \pm 0.02~{\rm mm}/\mu {\rm s} 
\label{eq:v_d}
\end{equation}
where the first correction to the drift length, applied to account for the field change in the gap, is based on the linear dependance of the drift velocity on the electric field within the current range and the second one ($\Delta\ell\simeq  2$ mm) for the thermal contraction at LAr T is inferred from NIST data on integrated coefficient of thermal expansion for the G10 material of the TPC structure.\\
The result of the drift velocity measurement in (\ref{eq:v_d}) refers to LAr in a temperature range of $\pm$ 0.1 K around the set point and under the operational electric field strength established with $\simeq 1$\% precision. 
The estimated relative error $\sigma_{\rm v}/$v$_d\simeq 2$\% includes uncertainties on the cathode and anode plane crossing time determination, related  to the intrinsic precision of the hit finding algorithm, and uncertainties on the effective drift length evaluation, mainly due to the error associated with the correction for the thermal contraction (not directly measured). 
The measured value is in very good agreement with expectations based on a common parametrization ~\cite{walkowiak} of the electron drift velocity  in liquid argon as a function of electric field strength, in the actual range of LAr temperature during the run, as reported in figure~\ref{drift-time} [Right].

\subsection{Electron lifetime}
\label{lifetime}

Accurate determination of the electron lifetime in LAr, an indicator of the level of chemical purity of the liquid, is a key 
element in the calorimetric energy reconstruction of ionization tracks and related particle identification. 
When measured, the electron lifetime allows to account for the free electron loss during the drift time due to attachment to electro-negative impurities (from material outgassing and residual leaks). \\
The impurities concentration in the liquid may vary in time, depending on the filters' removal efficiency and saturation level during operation in the recirculation/recondensation loop. The electron lifetime variation in time thus needs to be periodically monitored. A  $\tau_e$ measurement per DAQ-run (about 21 hrs on average) during the beam period was considered appropriate both for the determination of the charge correction factor to be used with the data collected during the corresponding DAQ-run and for filter saturation monitoring/replacement. This imposed the need for a fast, fully automated off-line procedure for the lifetime extraction during the physics run. \\
For this measurement, the abundant ``through-going track(s) event" sample (mainly muons from upstream $\nu$-interactions) is used. 
These impinge upon the front side of the detector and longitudinally cross the TPC volume, and 
have a narrow (and known) energy spectrum. Almost constant directionality and energy loss along the track make these tracks suitable for this measurement. \\
The accelerator beam-timing signal is used as the neutrino-event trigger for {\sf ArgoNeuT} and the through-going muon rate was found in the range of 60 (selected) events per hour. \\
Samples of about 1500 tracks per DAQ-run are used to extract the electron lifetime through a fully automated  off-line procedure derived from \cite{10m3}. 
Selected tracks span most (at least 120) of the wires in both Collection and Induction plane and are evenly distributed along the drift distance, from cathode to anode plane (see \cite{ArgoNeuT-mu-paper} for a detailed analysis of the through-going muon sample). 
The peak amplitude (i.e. the detected charge $Q_{det}$) of every hit of the tracks in the sample is plotted against its drift time. This is done separately for the hits of the Collection plane and the Induction plane.\\
\begin{figure}[tb]
\begin{centering}
\begin{tabular}{c c}
\hspace{-1.cm}
\includegraphics[height=3.5in]{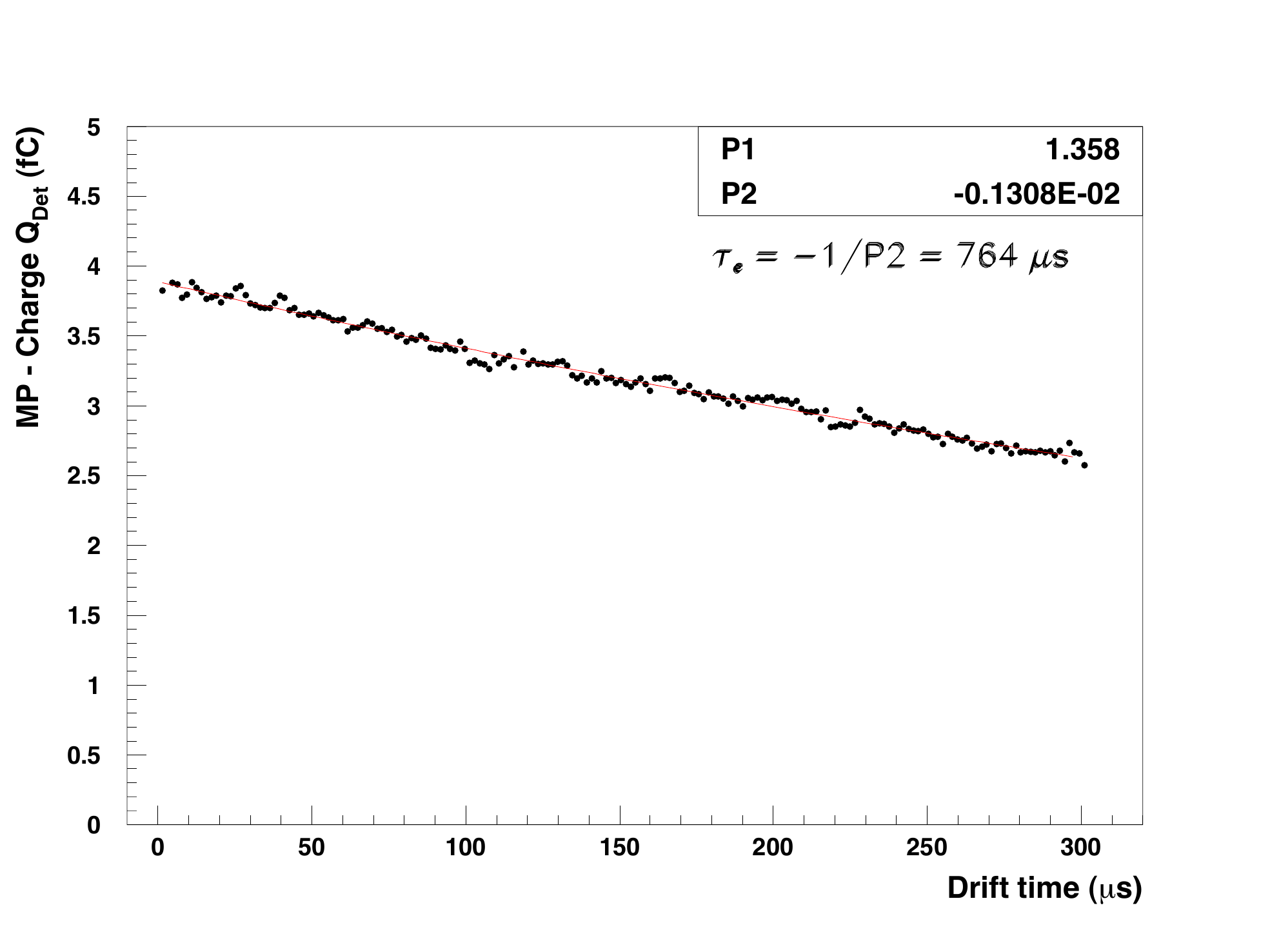}
\end{tabular}
\caption{The e-lifetime extraction from the  Collection plane with an exponential fit to the data. 
The fit gives an e-lifetime of 764$\pm\mathrm{3_{stat}}$~$\mu s$ (statistical error only) for the 
DAQ-run \#648 taken during the $\nu$-beam period.}
\label{lifetime_extract}
\end{centering}
\end{figure}
Each of the 2D scatter plots is broken up into time slices, eight samples wide (1.6~$\mu$s). Amplitudes of the hits in the slice are distributed in a 1D histogram for each time slice. The time slice width was chosen to be as small as possible, to minimize the smearing effects of a finite e-lifetime within the time region, while still allowing reasonable statistics per slice. The 1D histograms are fit using a convoluted Landau-Gaussian distribution. The Landau distribution describes the features of the energy loss by ionization and the Gaussian distribution accounts for fluctuations in the detected charge due to electronic noise, differences in track pitch length associated to the hit wire, 
electron diffusion and energy spectrum of the incident tracks. The fit has four parameters including the most probable value (MP) of the charge in the Landau distribution and the Gaussian spread ($\sigma_G$)  of the convoluted Gaussian function.
 A separate fit is performed for each of the 190 time slices of the total drift time from the cathode to the wire plane. \\
\begin{figure}[tb]
\begin{centering}
\begin{tabular}{c c}
\includegraphics[height=3.5in]{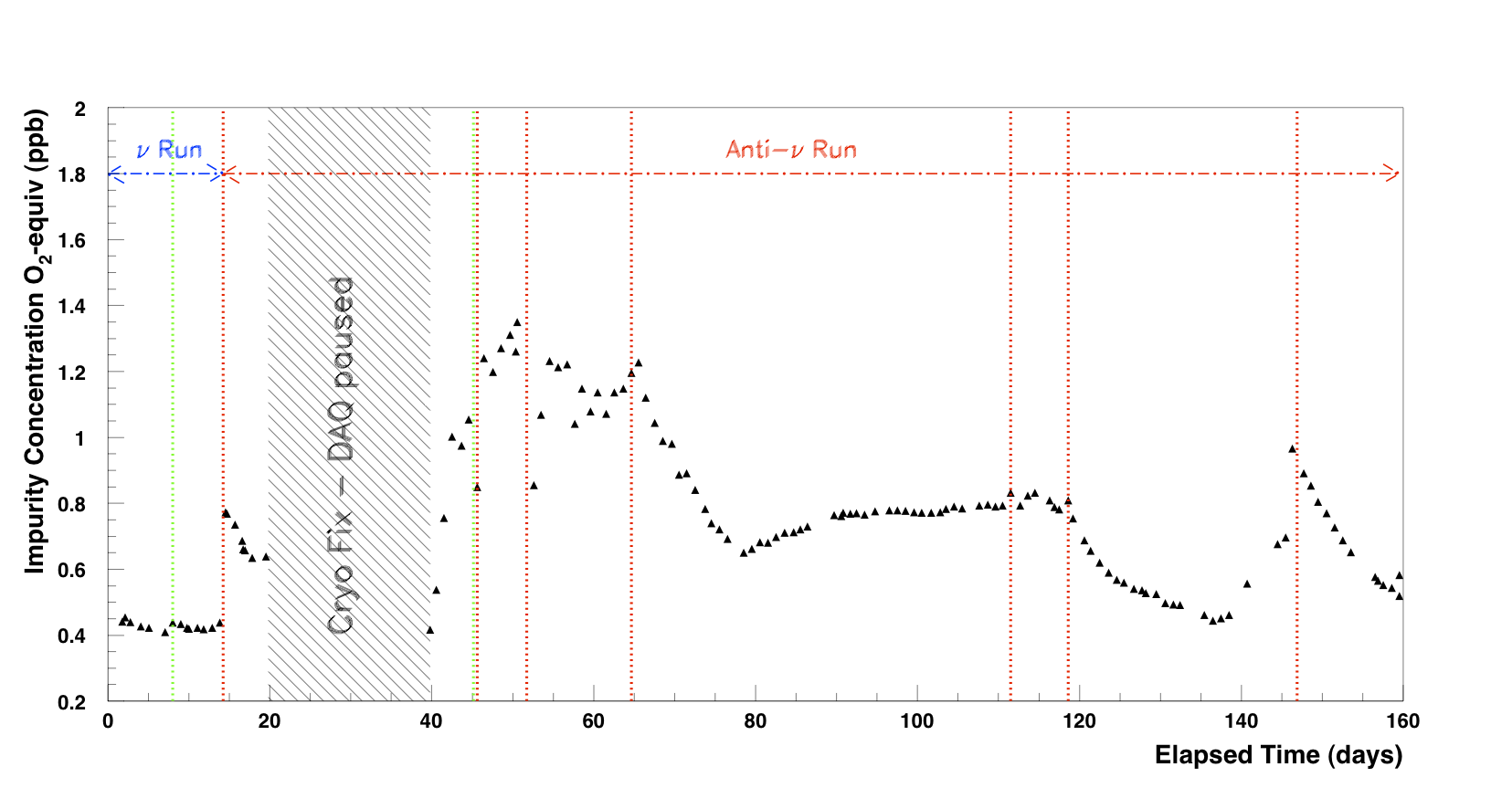}
\end{tabular}
\vspace{-0.6cm}
\caption{The O$_2$ equivalent impurity concentration in LAr (ppb) extracted from each DAQ-run as a function of elapsed time (days) since the start of the {\sf ArgoNeuT} physics run (about 160 days, beam in neutrino $\oplus$ antineutrino-mode). 
Vertical lines and band indicate hardware interventions on the purification/recirculation system: [red dotted] Filter exchange, [green dotted] 
GAr purge, [gray shaded] replacement of a cryocooler component - recirculation was halted during this period (about two weeks) and DAQ paused.}
\label{impurity_conc-run}
\end{centering}
\end{figure}
A plot of the MP charge vs drift time (time slice centre) for the Collection plane can be seen in figure~\ref{lifetime_extract}. The electron lifetime is extracted with an exponential function fit to the MP values for each plane.\\
The two independent measurements are combined to arrive at the e-lifetime value associated to the DAQ-run. \\
The e-lifetime measurement is minimally dependent on the time slice width. The extracted lifetime varies by less than one percent after choosing widths of 4, 8, 16, and 32 time samples.\\

From the e-lifetime the actual (O$_2$ equivalent) impurity concentration in LAr is inferred through Eq.\ref {eq:O2_conc}. 
The evolution of the impurity concentration (in ppb units) as a function of time for the entire run (spanning both neutrino and antineutrino-mode periods) is shown in figure~\ref{impurity_conc-run}.\\
Considering the maximum drift time of 300 $\mu$s in the LArTPC, a level of about 1 ppb was targeted as the
 maximum sustainable impurity concentration during operations. 
 The LAr purity level is maintained higher than this limit 
by the continuous operation of the GAr purification system (Sec.\ref{purification}).
The GAr recirculation however was unexpectedly halted for about two weeks, at the beginning of the anti-neutrino run [gray shaded  band in figure~\ref{impurity_conc-run}], due to a failure (and replacement) of a commercial component in the cryo-cooling system. 
The level of impurities increased soon after, as presumably due to diffusion of impurities from air leakage into the cryostat 
during cryogenics repairs. Purity was restored in about ten days of GAr recirculation through freshly regenerated filters.

\section{Conclusion}
\label{conclusion}
The {\sf ArgoNeuT} detector was successfully operated on the NuMI beam at Fermilab for an extended physics run period. All technical aspects of the detector were tested and performed satisfactorily, allowing for almost shift-free operation for the nine months long period  of the run. About 9000 neutrino events have been collected with excellent data quality, enabling 
physics studies and
software development towards high resolution imaging, accurate deposited energy reconstruction and powerful particle identification. 
A detailed description of the detector components, their characteristics and performance in working conditions were given, with the aim of providing a robust technical background information for future references to the {\sf ArgoNeuT} operation and physics outputs.
This information includes both hardware and software aspects of the experiment, from the cryogenic system implemented in the cooling and purification 
loop to the time projection chamber and associated read-out electronics, from the raw data filtering to the full 3D-imaging of the neutrino events.
Commissioning of the detector in the MINOS Near Detector Hall at Fermilab has been also reported as well as neutrino run details and display of the first neutrino events collected.  
A large sample of muons, generated by upstream neutrino interactions and crossing the {\sf ArgoNeuT} detector, have been collected. 
This sample was extremely useful as a tool for the characterization and monitoring of the main working parameters of the detector, lifetime and drift velocity  of the free electron charge in LAr, during the run period.
The main features of and the results from the analysis of this class of events were included as last topic of this report. 

The LArTPC detector technology as exploited with {\sf ArgoNeuT}
 provides exceptional capabilities, directly suited not only to the physics goals of the {\sf ArgoNeuT} experiment but also to
the broad goals of the ``Intensity Frontier" experiments in the current mid and long term scientific research planning in the US in general.
Signatures from processes of interest, such as inclusive $\nu_e$ interactions and exclusive proton decay channels, can thus 
be studied in great detail and in highly reduced background conditions.

\section{Acknowledgments}
We wish to acknowledge the support of the National Science Foundation, the Department
of Energy, and Fermilab in {\sf ArgoNeuT}Õs construction, operation, and data
analysis. In particular, 
we warmly thank the Fermilab Particle Physics Division technician crew led by J. Voirin for the invaluable contributions throughout the ArgoNeuT project, R. Schmitt and D. Markley for the essential contributions in the implementation of the cryogenic and process-monitoring systems, and  the technician crew based at the Fermilab Proton Assembly Building for the expert assistance during detector commissioning.
The {\sf ArgoNeuT} Collaboration acknowledges the cooperation of the MINOS Collaboration during the physics run 
and for providing their data for use in the analysis.

\end{document}